\documentclass[11pt,a4paper]{article}
\usepackage{amsmath}
\usepackage{amsfonts}
\usepackage{amssymb}
\usepackage{makeidx}
\usepackage{graphicx}
\usepackage[left=3cm,right=3cm,top=3cm,bottom=3cm]{geometry}
\usepackage{algorithm,algorithmic}

\usepackage{comment}

\newtheorem{theorem}{Theorem}

\newtheorem{result}[theorem]{Result}
\newtheorem{definition}[theorem]{Definition}

\newtheorem{remark}[theorem]{Remark}
\newtheorem{example}[theorem]{Example}

\newenvironment{proof}[1][Proof]{\noindent\textbf{#1.} }{\ \rule{0.5em}{0.5em}}

\begin{document}

\author{}
\title{\textbf{Robust estimation based on one-shot device test data under log-normal lifetimes}}
\date{}
\author{N. Balakrishnan\footnote{Department of Mathematics and Statistics, McMaster University, Hamilton, Ontario, Canada. ORCID: https://orcid.org/0000-0001-5842-8892}  \ and E. Castilla\footnote{Departamento de Matematica Aplicada,  Rey Juan Carlos University, Mostoles Campus,  28933  Madrid, Spain. ORCID:https://orcid.org/0000-0002-9626-6449.  email: elena.castilla@urjc.es.} }
\maketitle

\begin{abstract}
In this paper we present robust estimators for one-shot device test data under lognormal lifetimes. Based on these estimators, confidence intervals and Wald-type tests are also developed. Their robustness feature is illustrated through a simulation study and two numerical examples.
\end{abstract}

\noindent \textbf{Keywords:} Divergences; Reliability; Robustness, One-shot devices.\\
\noindent \textbf{Conflicts of interest:} All authors declare that they have no conflicts of interest.
\clearpage

\section{Introduction}

A destructive one-shot device is a product, system or weapon that can be used only once. After use, it gets destroyed or must be rebuilt. Some examples are munitions,  automobile air bags,  heat detectors and antigen tests.  The study of one-shot device data has been developed considerably recently, mainly motivated by the works of Ling (2012) and So (2016). For a complete review of all recent developments concerning the analysis of one-shot device data, one may refer to the  book by Balakrishnan et al. (2021).

Due to significant development in manufacturing technology, such one-shot devices usually possess long lifetimes, with a low  failure rate. That makes the initial data collection and processing stages more complex. To save time and cost, accelerated life tests (ALTs) are commonly employed as they induce early failures by testing items under high stresses, such as temperature, air-pressure or voltage. A very common type of ALT is the constant-stress ALT (CSALT),  which assumes that each device is subject to only pre-specified stress levels; see, for example, Balakrishnan and Ling (2014a) and Balakrishnan et al. (2022). Other forms of ALTs for one-shot devices have been also discussed in the literature; see Ling (2019), Ling and Hu (2020) and Zhu et al. (2021), among others.

From here on, we assume CSALTs are performed on $I$ groups of one-shot devices, each of which is subject to $J$ types of accelerating factors, and that in the $i$-th testing condition $K_{i}$ individuals are tested at a pre-specified inspection time $\tau_{i}$, for $i=1,\dots ,I$. Then, in the $i$-th test group, the number of failures, $n_i$, is  collected. The experimental condition and the data so obtained are as summarized in Table \ref{table:lognormal_model}. Note that one-shot device data are an extreme case of censoring since we do not observe any failure time, only the status of the device (failure or not) gets observed at the inspection time.

\begin{table}[htbp]  
\caption{Data collected on one-shot devices at multiple stress levels and collected at different inspection times. \label{table:lognormal_model}}
\center
\begin{tabular}{c ccc ccc}
\hline
 &  &   &   &  & Covariates &  \\ \cline{5-7}
Condition & Inspection Time & Devices & Failures & Stress $1$ & $\cdots $ & Stress $J$ \\ 
\hline
$1$ & $\tau_{1}$ & $K_{1}$ & $n_{1}$ & $x_{11}$ & $\cdots $ & $x_{1J}$ \\ 
$2$ & $\tau_{2}$ & $K_{2}$ & $n_{2}$ & $x_{21}$ & $\cdots $ & $x_{2J}$ \\ 
$\vdots $ & $\vdots $ & $\vdots $ & $\vdots $ & $\vdots $ &  & $\vdots $ \\ 
$I$ & $\tau_{I}$ & $K_{I}$ & $n_{I}$ & $x_{I1}$ & $\cdots $ & $x_{IJ}$ \\ 
\hline
\end{tabular}
\end{table}
Most of the existing works on one-shot devices are based on parametric models. Under the parametric model, the lifetimes are described by a statistical distribution which relates the stress factors to the model parameter vector $\boldsymbol{\theta}\in \Theta$. Inference for exponential, gamma and Weibull distributions has been widely developed for one-shot device test data in this context; see Balakrishnan and Ling (2012, 2013, 2014) and Balakrishnan et al. (2020a, 2020b), among others. However,  lognormal lifetime  distribution has not been studied in this set-up. While the hazard function (which measures the instantaneous rate of failure) for exponential distribution is always a constant, and that of Weibull and gamma are either increasing or decreasing, the lognormal distribution has increasing-decreasing behavior of hazard which is encountered often in practice as units usually experience early failure and then stabilize over time in terms of performance. Balakrishnan and Castilla (2022) developed an EM algorithm for the likelihood estimation of the model parameter vector based on one-shot device test data under lognormal distribution. However, the maximum likelihood estimator (MLE) is known for its lack of robustness. For this reason,  we develop here robust divergence-based inference for one-shot device test data under lognormal distribution, following an approach similar to that of  Balakrishnan et al. (2021).

This paper is organized as follows. Section \ref{sec:model} introduces the likelihood approach for one-shot device testing under the lognormal distribution. In Section \ref{sec:div}, we present the weighted minimum DPD estimators as an alternative to the MLEs. Their asymptotic distribution is  presented and a family of Wald-type test statistics is then developed. The robustness features of the new families of estimators and tests are evaluated by means of the boundedness of their Influence Function (IF) in Section \ref{sec:IF}. Section \ref{sec:num} presents the results of a detailed simulation study. In Section \ref{sec:ex} two examples are used to illustrate all the inferential results developed here. Finally, some concluding remarks are made in Section \ref{sec:CR}.

\section{Model description and likelihood inference \label{sec:model}}

We shall  assume that the lifetimes of the units, under the testing condition $i$, follow lognormal distribution with  probability density function and cumulative distribution function 
\begin{align}\label{eq:lognormal_f}
f(t;\boldsymbol{x}_i,\boldsymbol{\theta})=\frac{1}{\sqrt{2\pi}\sigma_i t}exp\left\{ -\frac{(\log(t)-\mu_i)^2}{2\sigma_i^2}\right\}, \quad t>0,
\end{align}
and
\begin{align}\label{eq:lognormal_F}
F(t;\boldsymbol{x}_i,\boldsymbol{\theta})=\Phi\left(\frac{\log(t)-\mu_i}{\sigma_i} \right), \quad t>0,
\end{align}
respectively. Here, $\Phi(\cdot)$ denotes the cumulative distribution function of the standard normal distribution, and  $\mu_i\in \mathbb{R}$ and $\sigma_i\in \mathbb{R}^+$ are, respectively, the scale and shape parameters, which we assume are related to the stress factors by
\begin{align*}%\label{eq:lognormal_parameters}
\mu_i=\sum_{j=0}^{J}a_jx_{ij} \quad \text{and} \quad
\sigma_i=\exp \left\{\sum_{j=0}^{J}b_jx_{ij} \right\},
\end{align*}
with $x_{i0}=1$ for all $i$. Here $\boldsymbol{\theta}=(a_0,\dots,a_J,b_0,\dots,b_J)^T\in \Theta=\mathbb{R}^{(2J+1)}$ is the model parameter vector. In this case, the mean lifetime and  the hazard function  are known to be (Johnson et al. (1994))

\begin{align}
E(\boldsymbol{x}_i,\boldsymbol{\theta})&=\exp \left\{\mu_i+\sigma_i^2/2 \right\},\label{eq:E}\\
h(t;\boldsymbol{x}_i,\boldsymbol{\theta})&=\frac{\frac{1}{\sqrt{2\pi}\sigma_i t}\exp\left\{ -\frac{(log( t)-\mu_i)^2}{2\sigma_i^2}\right\}}{1-\Phi\left(\frac{log( t)-\mu_i}{\sigma_i} \right)},\notag
\end{align}
respectively.

Instead of working with lognormal lifetimes, it is more convenient to work with log-lifetimes, $\omega_{ik}=log(t_{ik})$, since this belongs to a location-scale family of distributions; see Meeker (1984) and Yuan et al. (2018). In fact, the log-lifetimes follow a normal distribution with density and cumulative distribution functions as

\begin{align}
f_{\omega}(\omega;\boldsymbol{x}_i,\boldsymbol{\theta})&=\phi\left(\frac{\omega-\mu_i}{\sigma_i} \right)=\frac{1}{\sqrt{2\pi}\sigma_i}exp\left\{ -\frac{(\omega-\mu_i)^2}{2\sigma_i^2}\right\}, \quad -\infty<\omega<\infty, \notag\\
F_{\omega}(\omega;\boldsymbol{x}_i,\boldsymbol{\theta})&=\Phi\left(\frac{\omega-\mu_i}{\sigma_i} \right), \quad -\infty<\omega<\infty, \label{eq:Fw}
\end{align}
respectively. 

For convenience, let us now denote $\boldsymbol{z}=\left\{ W_i,K_i,n_i,i=1,\dots,I \right\}$ with $W_i=\log(\tau_i)$ for the observed data. Assuming independent observations, the  likelihood function based on the observed data is given by
\begin{align}\label{eq:lognormal_MLE}
\mathcal{L}(\boldsymbol{\theta};\boldsymbol{z})\propto \prod_{i=1}^I F^{n_i}_{\omega}(W_i;\boldsymbol{x}_i,\boldsymbol{\theta})R^{K_i-n_i}_{\omega}(W_i;\boldsymbol{x}_i,\boldsymbol{\theta}),
\end{align}
where $R_{\omega}(\omega;\boldsymbol{x}_i,\boldsymbol{\theta})=1-F_{\omega}(\omega;\boldsymbol{x}_i,\boldsymbol{\theta})$ is the reliability function at test condition $i$.  The corresponding MLE of $\boldsymbol{\theta}$, $\widehat{\theta}_{MLE}$,  will be obtained by maximization of (\ref{eq:lognormal_MLE}) or, equivalently, its logarithm,  as

\begin{equation}
 \widehat{\boldsymbol{\theta}}=\arg \max_{\boldsymbol{\theta}\in \Theta}\text{ }\log\mathcal{L}(\boldsymbol{\theta};\boldsymbol{z}).
 \label{eq:lognormal_MLE}
\end{equation}

Differentiating  (\ref{eq:lognormal_MLE}) with respect to $\boldsymbol{\theta}$ and equating to zero,  we have  the estimating equations of the MLE as

\begin{align*}
\sum_{i=1}^{I}\boldsymbol{\delta}_i\left( K_{i}F_{\omega}(W_i;\boldsymbol{x}_i,\boldsymbol{\theta})-n_{i}\right) \left( F_{\omega}^{ -1}(W_i;\boldsymbol{x}_i,\boldsymbol{\theta})+R_{\omega}^{ -1}(W_i;\boldsymbol{x}_i,\boldsymbol{\theta})\right) \boldsymbol{x}_i=\boldsymbol{0}_{2(J+1)},
%\label{eq:lognormal_est_eq}
\end{align*}
where   $\boldsymbol{0}_{2(J+1)}$ is the null column vector of dimension $2(J+1)$, 
\begin{equation}\label{eq:lognormal_Delta}
\boldsymbol{\delta}_i\boldsymbol{x}_i=\frac{\partial F_{\omega}(W_i;\boldsymbol{x}_i,\boldsymbol{\theta})}{\partial \boldsymbol{\theta}}=\left(\frac{1}{\sigma_i}\phi\left( \frac{W_i-\mu_i}{\sigma_i}\right), \frac{(W_i-\mu_i)}{\sigma_i}\phi\left( \frac{W_i-\mu_i}{\sigma_i}\right)
\right)^T\boldsymbol{x}_i,
\end{equation}
and $\phi(\cdot)$ is the density function of the standard normal distribution.

\subsection{Asymptotic distribution of the MLE and confidence intervals}\label{sec:MLE_asyCI}

The observed Fisher Information matrix for model parameters, $\boldsymbol{\mathcal{I}}_{obs}$, is given by the second-order derivatives of the observed log-likelihood function in (\ref{eq:lognormal_MLE}), given by

\begin{align*}
\boldsymbol{\mathcal{I}}_{obs}(\boldsymbol{\theta})=\sum_{i=1}^I K_{i} \left( F^{ -1}(IT_i;\boldsymbol{x}_i,\boldsymbol{\theta})+R^{ -1}(IT_i;\boldsymbol{x}_i,\boldsymbol{\theta})\right) \left(\frac{\partial  F(IT_i;\boldsymbol{x}_i,\boldsymbol{\theta})}{\partial \boldsymbol{\theta}} \right)\left(\frac{\partial  F(IT_i;\boldsymbol{x}_i,\boldsymbol{\theta})}{\partial \boldsymbol{\theta}^T} \right).
\end{align*}

It is well-known that the  asymptotic variance-covariance matrix of the MLEs of the model parameters can then be obtained by inverting the above observed Fisher information matrix as
\begin{align*}
\boldsymbol{V}(\boldsymbol{\theta})=\boldsymbol{\mathcal{I}}_{obs}^{-1}(\boldsymbol{\theta}).
\end{align*}
In particular, the variance-covariance matrix  is the Rao-Cramer lower bound, implying the efficiency of the MLEs if the model is correctly specified.

The asymptotic variance of the MLE of lifetime characteristics, such as reliability and mean lifetime of units at normal operating conditions $(\omega_0,\boldsymbol{x}_0)$, can be computed readily by employing delta method, as

\begin{align*}
\boldsymbol{V}_R(\boldsymbol{\theta})&=\boldsymbol{P}^T_R(\widehat{\boldsymbol{\theta}})\boldsymbol{V}(\boldsymbol{\theta})\boldsymbol{P}_R(\widehat{\boldsymbol{\theta}}),\\
\boldsymbol{V}_E(\boldsymbol{\theta})&=\boldsymbol{P}^T_E(\widehat{\boldsymbol{\theta}})\boldsymbol{V}(\boldsymbol{\theta})\boldsymbol{P}_E(\widehat{\boldsymbol{\theta}}),
\end{align*}
where

\begin{align*}
\boldsymbol{P}_R(\widehat{\boldsymbol{\theta}})=\left. \frac{\partial R_{\omega}(\omega_0;\boldsymbol{x}_0,\boldsymbol{\theta})}{\partial \boldsymbol{\theta}} \right|_{\boldsymbol{\theta}=\widehat{\boldsymbol{\theta}}} \quad \text{and} \quad \boldsymbol{P}_E(\widehat{\boldsymbol{\theta}})=\left. \frac{\partial E_{\omega}(\boldsymbol{x}_0,\boldsymbol{\theta})}{\partial \boldsymbol{\theta}} \right|_{\boldsymbol{\theta}=\widehat{\boldsymbol{\theta}}}.
\end{align*}
In practice, as we do not know the real value of $\boldsymbol{\theta}$, we will need to replace it by the estimated $\widehat{\boldsymbol{\theta}}$. 
  
The $100(1-\alpha)\%$ asymptotic confidence interval for any parameter of interest $\upsilon$ is  given by
\begin{align*}
(\widehat{\upsilon}-z_{1-\alpha/2}se(\widehat{\upsilon}),\widehat{\upsilon}+z_{1-\alpha/2}se(\widehat{\upsilon})),
\end{align*}
where $z_{1-\alpha/2}$ is the $1-\alpha/2$ standard normal quantile and $se(\widehat{\upsilon})$ is the standard error of $\widehat{\upsilon}$ obtained  from the asymptotic variance-covariance matrix. However, the asymptotic confidence interval is based on the asymptotic properties of the MLE, and so it may be satisfactory only for large sample sizes. Further, we would need to truncate the bounds of the confidence intervals for the mean lifetime and reliability of the devices, since the mean lifetime has to be positive and, similarly, the reliability lies between 0 and 1. For this reason, Balakrishnan and Castilla (2022) studied the behaviour of  the hyperbolic arcsecant (arsech) and logit transformations of confidence intervals of reliability under the lognormal distribution.  

In the arsech-approach, the $100(1-\alpha)\%$ asymptotic confidence interval for reliability takes on the form 
\begin{align}
\left(\frac{2}{\exp(-U_f)+\exp(U_f)}, \frac{2}{\exp(-L_f)+\exp(L_f)} \right),
\label{eq:lognormal_arsechlogit}
\end{align}
with
\begin{align*}
U_f&=\widehat{f}+z_{1-\frac{\alpha}{2}}se(\widehat{f}),\quad L_f=\widehat{f}-z_{1-\frac{\alpha}{2}}se(\widehat{f}),
\end{align*}
and
\begin{align*}
\widehat{f}=\log \left( \frac{1+\sqrt{1-\widehat{R}^2}}{\widehat{R}}\right), \quad se(\widehat{f})=\frac{se(\widehat{R})}{\widehat{R}\sqrt{1-\widehat{R}^2}},
\end{align*}
where $\widehat{R}$ denotes, for simplicity, the estimated reliability. 
In the logit-approach, the $100(1-\alpha)\%$ asymptotic confidence interval for reliability is of the form 
\begin{align}
\left(\frac{\widehat{R}}{\widehat{R}+(1-\widehat{R})S}, \frac{\widehat{R}}{\widehat{R}+(1-\widehat{R})/S}\right),
\label{eq:lognormal_cilogit}
\end{align}
where $S=\exp\left(z_{1-\frac{\alpha}{2}}\frac{se(\widehat{R})}{\widehat{R}(1-\widehat{R})} \right)$. For more details concerning the confidence intervals in  (\ref{eq:lognormal_arsechlogit}) and (\ref{eq:lognormal_cilogit}), see Balakrishnan and Ling (2014b). These authors also studied the performance of the log-approach for the construction of a $100(1 - \alpha)\%$ confidence interval for the expected lifetime, $\widehat{T}$,  of the form
\begin{align}
\left(\widehat{T} \exp\left\{-\frac{z_{1-\frac{\alpha}{2}}se(\widehat{T})}{\widehat{T}} \right\},\widehat{T} \exp\left\{\frac{z_{1-\frac{\alpha}{2}}se(\widehat{T})}{\widehat{T}} \right\} \right).
\end{align}

\subsection{Wald test}

The Wald test is a well-known  multivariate test of hypothesis that allows testing a set of
parameters of the considered model simultaneously.

Let us consider the function $\boldsymbol{m:}\mathbb{R}^{2(J+1)}\longrightarrow\mathbb{R}^{r}$, where $r\leq 2(J+1)$. Then, $\boldsymbol{m}\left( \boldsymbol{\theta }\right) =\boldsymbol{0}_{r}$ represents a composite null hypothesis. We assume that the $2(J+1)\times r$ matrix 
\[
\boldsymbol{M}\left( \boldsymbol{\theta }\right) =\frac{\partial \boldsymbol{m}^{T}\left( \boldsymbol{\theta }\right) }{\partial \boldsymbol{\theta }}
\]%
exists and is continuous in $\boldsymbol{\theta }$, with rank $\boldsymbol{M}\left( \boldsymbol{\theta }\right) =r$. Then, for testing 
\begin{equation}
H_{0}:\boldsymbol{\theta} \in {\Theta }_{0}\text{ against }H_{1}:\boldsymbol{\theta} \notin {\Theta }_{0},  \label{eq:nullhyp}
\end{equation}%
with ${\Theta }_{0}=\left\{ \boldsymbol{\theta }\in \Theta:\boldsymbol{m}\left( \boldsymbol{\theta }\right) =\boldsymbol{0}_{r}\right\}$, we can consider the following Wald test statistic:
\begin{equation}
W_{K}(\widehat{\boldsymbol{\theta }})=K\boldsymbol{m}^{T}(\widehat{\boldsymbol{\theta }})\left( \boldsymbol{M}^{T}(\widehat{\boldsymbol{\theta }})\boldsymbol{\Sigma }(\widehat{\boldsymbol{\theta }})\boldsymbol{M}(\widehat{\boldsymbol{\theta }})\right) ^{-1}\boldsymbol{m}(\widehat{\boldsymbol{\theta }}),
\label{eq:Wald}
\end{equation}%
where, for further convenience,  $\boldsymbol{\Sigma }(\widehat{\boldsymbol{\theta }})=\frac{\boldsymbol{V }(\widehat{\boldsymbol{\theta }})}{K}$. Here, $K=\sum_{i=1}^IK_i$ is the total number of devices.

The asymptotic null distribution of the  Wald test statistic, given in (\ref{eq:Wald}), is a chi-squared ($\chi ^{2}$) distribution with $r$ degrees of freedom, i.e., 

\[
W_{K}(\widehat{\boldsymbol{\theta }}_)\underset{K\rightarrow \infty }{\overset{\mathcal{L}}{\longrightarrow }}\chi _{r}^{2}.
\]
Based on this, we will reject the null hypothesis in (\ref{eq:nullhyp}) if 
\begin{equation*}
W_{K}(\widehat{\boldsymbol{\theta }})>\chi _{r,\alpha }^{2},
\end{equation*}%
where $\chi _{r,\alpha }^{2}$ is the upper $\alpha$ percentage point of $\chi _{r}^{2}$ distribution.

\begin{example}
In the context of one-shot device testing under multiple stress factors,  we may be interesting in checking whether there is a significant relationship between  the $j-$th stress factor and the device lifetime. This can be tested through  Wald test in (\ref{eq:Wald}), where   $\boldsymbol{m}\left( \boldsymbol{\theta}\right)=(\theta_j,\theta_{(J+1)+j})$, and

$$
\boldsymbol{M}^T\left( \boldsymbol{\theta}\right)=
\begin{pmatrix}
0,&\dotsc,&\overset{(j+1)}{1},& \dotsc,& \overset{(J+1)}{0}\hspace{-0.17cm},&\dotsc,&\overset{(J+1+j)}{0},& \dotsc,& \overset{(2J+2)}{0}\\
0,&\dotsc,&0,& \dotsc,& 0\hspace{-0.17cm},&\dotsc, & 1,& \dotsc,& 0
\end{pmatrix}.
$$
\end{example}

%\clearpage
\section{Divergence-based inference \label{sec:div}}
We first introduce some notation in order to define the MLE on the basis of  Kullback-Leibler divergence. For each testing condition $i$, $i=1,\dots,I$, the empirical and theoretical probability vectors are given, respectively, by
\begin{align}
\widehat{\boldsymbol{p}}_{i}&=\left( \widehat{p}_{i1},\widehat{p}_{i2}\right)^{T}=\left( \frac{n_{i}}{K_{i}},1-\frac{n_{i}}{K_{i}}\right)^{T}, \label{eq:Mult_emp_prob_vector}\\
\boldsymbol{\pi }_{i}(\boldsymbol{\theta})&=(\pi _{i1}(\boldsymbol{\theta}),\pi _{i2}(\boldsymbol{\theta}))^{T}=(F_{\omega}(W_{i};\boldsymbol{x}_i,\boldsymbol{\theta}),R_{\omega}(W_{i};\boldsymbol{x}_i,\boldsymbol{\theta}))^T.  \label{eq:Mult_theo_prob_vector}
\end{align}%
The empirical and theoretical vectors  provide, for each testing condition, the probability of failure and success based  on the data collected and on the theoretical parametric model considered, respectively.

Let us consider the weighted Kullback-Leibler divergence measure of all the units given by
\begin{align}
\sum_{i=1}^{I}\frac{K_{i}}{K}d_{KL}(\widehat{\boldsymbol{p}}_{i},\boldsymbol{\pi }_{i}(\boldsymbol{\theta}))=\frac{1}{K}\sum_{i=1}^{I}K_i\left[\widehat{p}_{i1}\log \left( \dfrac{\widehat{p}_{i1}}{\pi _{i1}(\boldsymbol{\theta})}\right) + \widehat{p}_{i2}\log \left( \dfrac{\widehat{p}_{i2}}{\pi _{i2}(\boldsymbol{\theta})}\right)\right], \label{eq:lognormal_dkull}
\end{align} 
where $K=\sum_{i=1}^IK_i$ is the total number of devices.\\

The following result establishes the relationship between the weighted Kullback-Leibler divergence in  (\ref{eq:lognormal_dkull}) and the MLE. 

\begin{result}\label{th:KL_MLE}
The MLE can be obtained as the minimization of the weighted Kullback-Leibler divergence measure between probability vectors in (\ref{eq:Mult_emp_prob_vector}) and (\ref{eq:Mult_theo_prob_vector}); that is,
\begin{equation}
\widehat{\boldsymbol{\theta}}=\underset{\boldsymbol{\theta}\in \Theta }{\arg \min }\sum_{i=1}^{I}\frac{K_{i}}{K}d_{KL}(\widehat{\boldsymbol{p}}_{i},\boldsymbol{\pi }_{i}(\boldsymbol{\theta})).  \label{eq:lognormal_mle_kull}
\end{equation}
\end{result}

\begin{remark}
Result  \ref{th:KL_MLE} provides us the basis for the development of alternative estimators for one-shot device testing. The idea is to choose a different divergence measure (between empirical and theoretical probability vectors) to be minimized. In particular, the DPD is well-known for its robustness properties.
\end{remark}

Given the probability vectors $\widehat{\boldsymbol{p}}_{i}$ and $\boldsymbol{\pi }_{i}(\boldsymbol{\theta})$ defined in (\ref{eq:Mult_emp_prob_vector}) and  (\ref{eq:Mult_theo_prob_vector}), respectively, the density power divergence (DPD) between the two probability vectors (see  Basu et al. (1998)), with tuning parameter $\beta\geq0$, is given by 
\begin{align}
d_{\beta }(\widehat{\boldsymbol{p}}_{i},\boldsymbol{\pi }_{i}(\boldsymbol{\theta})) &=\left( \pi _{i1}^{\beta +1}(\boldsymbol{\theta})+\pi _{i2}^{\beta +1}(\boldsymbol{\theta})\right) -\frac{\beta +1}{\beta }\left( \widehat{p}_{i1}\pi_{i1}^{\beta }(\boldsymbol{\theta})+\widehat{p}_{i2}\pi _{i2}^{\beta }(\boldsymbol{\theta})\right)  \notag \\
& \hspace{4cm}+\frac{1}{\beta }\left( \widehat{p}_{i1}^{\beta +1}+\widehat{p}_{i2}^{\beta +1}\right) ,\quad \text{if }\beta >0,  \label{eq:lognormal_DPD_long}\\
d_{\beta =0}(\widehat{\boldsymbol{p}}_{i},\boldsymbol{\pi }_{i}(\boldsymbol{\theta}))&=\lim_{\beta \rightarrow 0^{+}}d_{\beta }(\widehat{\boldsymbol{p}}_{i},\boldsymbol{\pi }_{i}(\boldsymbol{\theta}))=d_{KL}(\widehat{\boldsymbol{p}}_{i},\boldsymbol{\pi }_{i}(\boldsymbol{\theta})).
\end{align}%
As  the term $\frac{1}{\beta }\left( \widehat{p}_{i1}^{\beta +1}+\widehat{p}_{i2}^{\beta +1}\right) $ does not have any role in the minimization with respect to $\boldsymbol{\theta}$ in (\ref{eq:lognormal_DPD_long}), we can consider the equivalent measure for $\beta>0$ as
\begin{equation}
d_{\beta }^{\ast }(\widehat{\boldsymbol{p}}_{i},\boldsymbol{\pi }_{i}(\boldsymbol{\theta}))=\left( \pi _{i1}^{\beta +1}(\boldsymbol{a})+\pi_{i2}^{\beta +1}(\boldsymbol{\theta})\right) -\frac{\beta +1}{\beta }\left( \widehat{p}_{i1}\pi _{i1}^{\beta }(\boldsymbol{\theta})+\widehat{p}_{i2}\pi_{i2}^{\beta }(\boldsymbol{\theta})\right) .  \label{eq:lognormal_DPD_short1}
\end{equation}

Based on Result \ref{th:KL_MLE}, we can now define the weighted minimum DPD estimators for the one-shot device model with multiple stress levels.
\begin{definition}
Let us consider the framework in Table \ref{table:lognormal_model}, and define the weighted minimum DPD estimator for $\boldsymbol{\theta}$ as
\begin{equation*}
\widehat{\boldsymbol{\theta}}_{\beta }=\underset{\boldsymbol{\theta}\in \Theta}{\arg\min }\sum_{i=1}^{I}\frac{K_{i}}{K}d_{\beta }^{\ast }(\widehat{\boldsymbol{p}}_{i},\boldsymbol{\pi }_{i}(\boldsymbol{\theta})),\quad \text{for }\beta >0,
\end{equation*}
where $d_{\beta }^{\ast }(\widehat{\boldsymbol{p}}_{i},\boldsymbol{\pi }_{i}(\boldsymbol{\theta}))$  is as given in (\ref{eq:lognormal_DPD_short1}), and $\widehat{\boldsymbol{p}}_{i}$ and $\boldsymbol{\pi }_{i}(\boldsymbol{\theta})$ are as given in (\ref{eq:Mult_emp_prob_vector}) and (\ref{eq:Mult_theo_prob_vector}), respectively. For $\beta =0$, we have the MLE, $\widehat{\boldsymbol{\theta}}$.
\end{definition}

Differentiating $d_{\beta }^{\ast }(\widehat{\boldsymbol{p}}_{i},\boldsymbol{\pi }_{i}(\boldsymbol{\theta}))$ with respect to $\boldsymbol{\theta}$ and equating to zero, we have  for $\beta\geq0$, the estimating system of equations to be
\begin{align*}
&\sum_{i=1}^{I} \boldsymbol{\delta}_i \left(  K_{i}F_{\omega}(W_i;\boldsymbol{\theta}) -n_{i}\right)   \left[F^{\beta-1}_{\omega}(W_i;\boldsymbol{\theta})  + R^{\beta-1}_{\omega}(W_i;\boldsymbol{\theta})   \right]\boldsymbol{x}_i =\boldsymbol{0}_{2(J+1)},
\end{align*}
with $\boldsymbol{\delta}_i$ as in (\ref{eq:lognormal_Delta}).

A detailed review of divergence-based robust inferential methods for one-shot device testing under different lifetime distributions can be found in Balakrishnan et al. (2021). We now derive the asymptotic distribution of he proposed weighted minimum DPD estimators, which will facilitate the development of a new family of Wald-type tests for hypothesis testing.

\subsection{Asymptotic distribution and confidence intervals}

In the following result, we present the asymptotic distribution of the proposed weighted minimum DPD estimators. 

\begin{result}\label{th:lognormal_asymp}
Let $\boldsymbol{\theta}^0$ be the true value of the parameter $\boldsymbol{\theta}$. Then the asymptotic distribution of the weighted minimum DPD estimator $\widehat{\boldsymbol{\theta}}_{\beta}$ is given by
 \begin{equation*}
\sqrt{K}\left( \widehat{\boldsymbol{\theta}}_{\beta }-\boldsymbol{\theta}^{0}\right) 
\overset{\mathcal{L}}{\underset{K\mathcal{\rightarrow }\infty }{\longrightarrow }}\mathcal{N}\left( \boldsymbol{0}_{2(J+1)},\boldsymbol{{\Sigma}}_{\beta }(\boldsymbol{\theta}^{0})\right) ,
\end{equation*}%
where 
\begin{align}\label{eq:Sigma}
\boldsymbol{{\Sigma}}_{\beta }(\boldsymbol{\theta})=K\boldsymbol{{V}}_{\beta }(\boldsymbol{\theta})=\boldsymbol{{J}}_{\beta }^{-1}(\boldsymbol{\theta})\boldsymbol{{K}}_{\beta }(\boldsymbol{\theta})\boldsymbol{{J}}_{\beta }^{-1}(\boldsymbol{\theta}),
\end{align}
with
\begin{align*}
\boldsymbol{{J}}_{\beta }(\boldsymbol{\theta})& =\sum_{i}^{I}\frac{K_i}{K}\boldsymbol{\Delta}_{i}\left( F_{\omega}^{\beta -1}(W_{i};\boldsymbol{x}_i,\boldsymbol{\theta})+R_{\omega}^{\beta -1}(W_{i};\boldsymbol{x}_i,\boldsymbol{\theta})\right)\boldsymbol{x}_{i}\boldsymbol{x}_{i}^{T} , \\
\boldsymbol{{K}}_{\beta }(\boldsymbol{\theta})& =\sum_{i}^{I}\frac{K_i}{K}\boldsymbol{\Delta}_{i}F_{\omega}(W_{i};\boldsymbol{x}_i,\boldsymbol{\theta})R_{\omega}(W_{i};\boldsymbol{x}_i,\boldsymbol{\theta})  \left( F_{\omega}^{\beta -1}(W_{i};\boldsymbol{x}_i,\boldsymbol{\theta})+R_{\omega}^{\beta -1}(W_{i};\boldsymbol{x}_i,\boldsymbol{\theta})\right)^2\boldsymbol{x}_{i}\boldsymbol{x}_{i}^{T},
\end{align*}
and $\boldsymbol{\Delta}_i=\boldsymbol{\delta}_i\boldsymbol{\delta}_i^T$.
\end{result}

The proof of this result follows  from Theorem 3.1 of Ghosh and Basu (2013); see Appendix \ref{App} for more details. Using the asymptotic distribution of the weighted minimum DPD estimator, given in Result \ref{th:lognormal_asymp}, we can construct  confidence intervals similar to those detailed in Section \ref{sec:MLE_asyCI} for the model parameters.

\subsection{Wald-type tests}

Let us consider again the problem of testing the null hypothesis  in (\ref{eq:nullhyp}).  We can consider the following Wald-type test statistics as a generalization of the classical Wald test defined in (\ref{eq:Wald}):
\begin{equation}\label{eq:Waldeq}
W_{K}( \widehat{\boldsymbol{\theta}}_{\beta }) =K\boldsymbol{m}^{T}( \widehat{\boldsymbol{\theta}}_{\beta }) \left( \boldsymbol{M}^{T}( \widehat{\boldsymbol{\theta}}_{\beta }) \boldsymbol{\Sigma }( \widehat{\boldsymbol{\theta}}_{\beta }) \boldsymbol{M}( \widehat{\boldsymbol{\theta}}_{\beta }) \right) ^{-1}\boldsymbol{m}( \widehat{\boldsymbol{\theta}}_{\beta }) ,
\end{equation}%
where $\boldsymbol{\Sigma }( \boldsymbol{\theta}) $ is as given in (\ref{eq:Sigma}). The classical Wald test given in (\ref{eq:Wald}) is then deduced for $\beta=0$.

\begin{result}
\label{th:asymp_test} The asymptotic null distribution of the proposed Wald-type test statistics, given in  (\ref{eq:Waldeq}), is a chi-squared ($\chi^2$) distribution with $r$ degrees of freedom; that is,

\begin{equation*}
W_{K}( \widehat{\boldsymbol{\theta}}_{\beta }) \underset{K\rightarrow\infty }{\overset{\mathcal{L}}{\longrightarrow }}\chi _{r}^{2}.
\end{equation*}
\end{result}

Result \ref{th:asymp_test} shows that we can reject the null hypothesis in (\ref{eq:nullhyp}) if 
\begin{equation}
\label{eq:asymp_test_beta}
W_{K}( \widehat{\boldsymbol{\theta}}_{\beta }) >\chi _{r,\alpha }^{2},
\end{equation}
where $\chi _{r,\alpha }^{2}$ is the upper $\alpha$ percentage point  of $\chi _{r }^{2}$ distribution; see Appendix \ref{App} for  details on the proof of Result \ref{th:asymp_test}.

%\clearpage

\section{Study of the Influence Function \label{sec:IF}}

The Influence Function (IF) is an important concept  in   statistical robustness theory. Introduced by Hampel et al. (1986). The (first-order) IF of an estimator, as a function of $t$, measures the standardized asymptotic bias (in its first-order approximation) caused by the infinitesimal contamination at the point $t$. Larger the value of $t$ is, the less robust  the estimator is. 

Let us denote by ${G}_{i}$ the true distribution function of a Bernoulli random variable with an unknown probability of success, for the $i$-th group of $K_{i}$ observations, having mass function ${g}_{i}$. Similarly, by ${F}_{i,\boldsymbol{\theta}}$ the distribution function of a Bernoulli random variable having a probability of success as $\pi_{i1}(\boldsymbol{\theta})$, with probability mass function ${f}_{i}(\cdot,\boldsymbol{\theta})$ ($i=1,...,I$), which is related to the considered model. In vector notation, we consider $\boldsymbol{G}=({G}_{1}\otimes\boldsymbol{1}_{K_{1}}^{T},\dots,{G}_{I}\otimes\boldsymbol{1}_{K_{I}}^{T})^{T}$ and $\boldsymbol{F}_{\boldsymbol{\theta}}=({F}_{1,\boldsymbol{\theta}}\otimes\boldsymbol{1}_{K_{1}}^{T},\dots,{F}_{I,\boldsymbol{\theta}}\otimes\boldsymbol{1}_{K_{I}}^{T})^{T}$.\\

For any estimator defined in terms of a statistical functional $\boldsymbol{U}(\boldsymbol{G})$ in the set-up of data from the true distribution function $\boldsymbol{G}$, its IF  is defined as 
\begin{align*}
{IF}(\boldsymbol{t},\boldsymbol{U},\boldsymbol{G})=\lim_{\varepsilon\downarrow0}\frac{\boldsymbol{U}(\boldsymbol{G}_{\varepsilon,\boldsymbol{t}})-\boldsymbol{U}(\boldsymbol{G})}{\varepsilon}=\left.  \frac{\partial\boldsymbol{U}(\boldsymbol{G}_{\varepsilon,\boldsymbol{t}})}{\partial\varepsilon}\right\vert _{\varepsilon=0^{+}},
\end{align*}
where $\boldsymbol{G}_{\varepsilon,\boldsymbol{t}}=(1-\varepsilon)\boldsymbol{G}+\varepsilon\Delta_{\boldsymbol{t}}$, with $\varepsilon$ being the contamination proportion and $\Delta_{\boldsymbol{t}}$ being the distribution function of the degenerate random variable at the contamination point 
$$\boldsymbol{t}=(t_{11},...,t_{1K_{1}},...,t_{I1},...,t_{IK_{I}})^{T}\in \mathbb{R}^{I K}.$$
We first need to define the statistical functional $\boldsymbol{U}_{\beta}(\boldsymbol{G})$ corresponding to the weighted minimum DPD estimator as the minimizer of the weighted sum of DPDs between the true and model densities. This is defined as the minimizer of

\begin{align}\label{eq:robust_Hmin}
\sum_{i=1}^I \frac{K_{i}}{K}\left\{ \sum_{y \in \{ 0,1\}}\left[f_{i}^{\beta+1}(y,\boldsymbol{\theta})-\frac{\beta+1}{\beta}f_{i}^{\beta}(y,\boldsymbol{\theta})g_{i}(y) \right]\right\},
\end{align}
where $g_{i}(y)$ is the probability mass function associated to $G_{i}$ and 
\begin{align*}
f_{i}(y,\boldsymbol{\theta})=y\pi_{i1}(\boldsymbol{\theta})+(1-y)\pi_{i2}(\boldsymbol{\theta}), \quad y \in \{0,1 \}.
\end{align*}
If we choose  $g_{i}(y)\equiv f_{i}(y,\boldsymbol{\theta})$, the expression in  (\ref{eq:robust_Hmin}) is minimized at $\boldsymbol{\theta}=\boldsymbol{\theta}^0$, implying the Fisher consistency of the minimum DPD estimator functional $\boldsymbol{U}_{\beta}(\boldsymbol{G})$ in the considered model.\\

We can derive the IF of the minimum DPD estimators at $F_{\boldsymbol{\theta}^0}$ with respect to the $k$-th observation of the $i_0$-th group and with respect to all the observations as in the following results:

\begin{result}\label{th:IF1}
Let us consider the one-shot device testing under the lognormal distribution. Then, the IF with respect to the $k$-th observation of the $i_0$-th group is given by
\begin{align}  \label{eq:robust_IF_FINAL}
IF(t_{i_0,k},\boldsymbol{U}_{\beta},F_{\boldsymbol{\theta}^0})=&\boldsymbol{J}^{-1}_{\beta}(\boldsymbol{\theta}^0)\frac{K_{i_0}}{K} \boldsymbol{\delta}_{i_0} \boldsymbol{x}_{i_0}  \\
&\times \left(F_{\omega}^{\beta-1}(W_{i_0};\boldsymbol{\theta}^0)+R_{\omega}^{\beta-1}(W_{i_0};\boldsymbol{\theta}^0) \right) \left( F_{\omega}(W_{i_0};\boldsymbol{\theta}^0)-\Delta^{(1)}_{t_{i_0,k}}\right),  \nonumber
\end{align}
where $\Delta^{(1)}_{t_{i_0},k}$ is the degenerating function at point $(t_{i_0},k)$.\\
\end{result}

\begin{result}\label{th:IF2}
Let us consider the one-shot device testing under the lognormal distribution. Then, the IF with respect to all the observations is given by
\begin{align}  \label{eq:robust_IF_FINAL_TOTAL}
IF(\boldsymbol{t},\boldsymbol{U}_{\beta},F_{\boldsymbol{\theta}^0})=&\boldsymbol{J}^{-1}_{\beta}(\boldsymbol{\theta}^0)\sum_{i=1}^I\frac{K_{i}}{K} \boldsymbol{\delta}_{i} \boldsymbol{x}_{i}  \\
&\times \left(F_{\omega}^{\beta-1}(W_{i};\boldsymbol{\theta}^0)+R_{\omega}^{\beta-1}(W_{i}; \boldsymbol{\theta}^0) \right) \left( F_{\omega}(W_{i};\boldsymbol{\theta}^0)-\Delta^{(1)}_{t_{i}}\right),  \nonumber
\end{align}
where $\Delta^{(1)}_{t_{i}}=\sum_{k=1}^{K_i}\Delta^{(1)}_{t_{i,k}}$.
\end{result}

Proofs of Result \ref{th:IF1} and Result \ref{th:IF2} require some heavy manipulations. We omit them here, but one may refer to Balakrishnan et al. (2020b) for a similar derivation for the case of the Weibull distribution. 

\begin{figure}[p]
\centering
\begin{tabular}{cc}
	\includegraphics[scale=0.4]{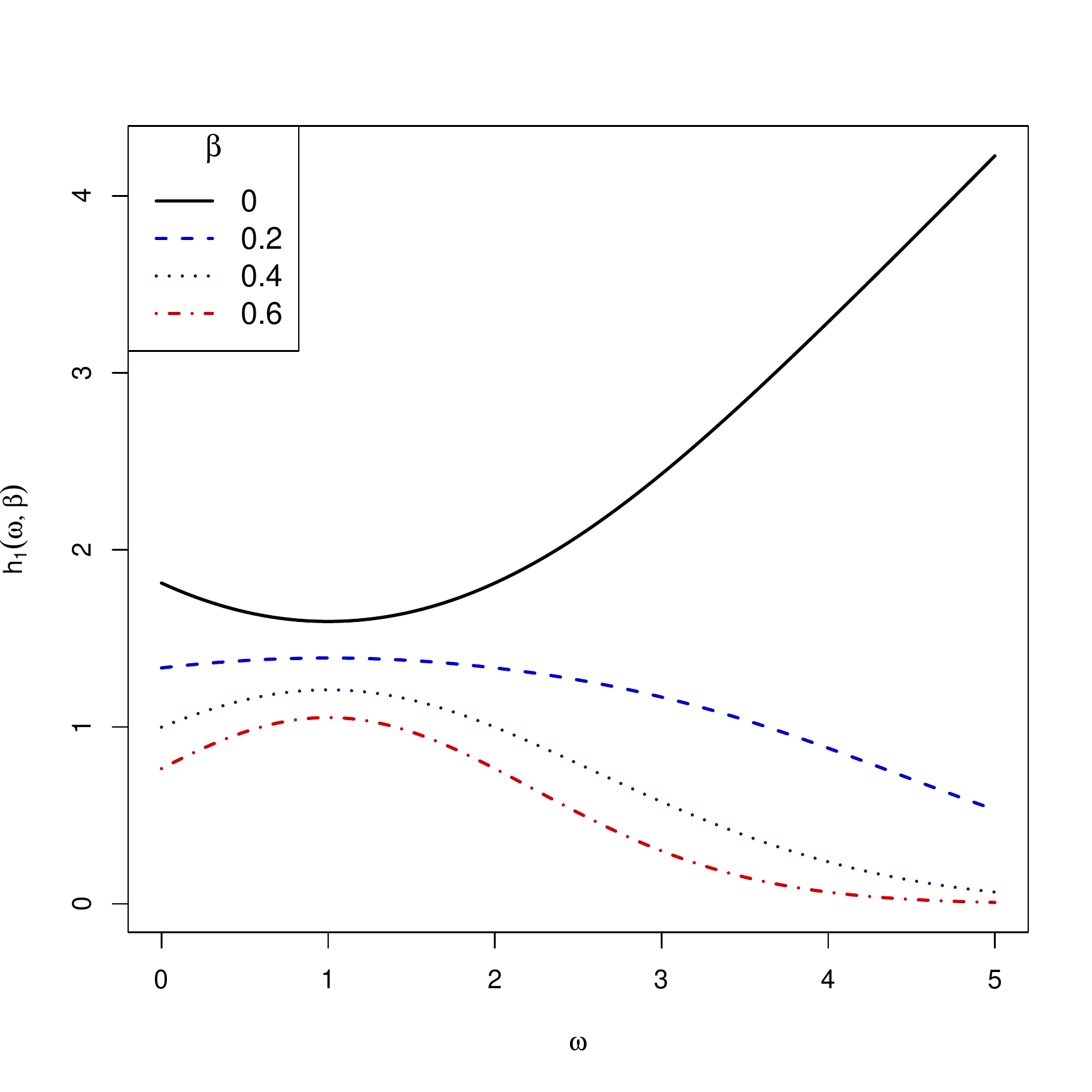}&
	\includegraphics[scale=0.4]{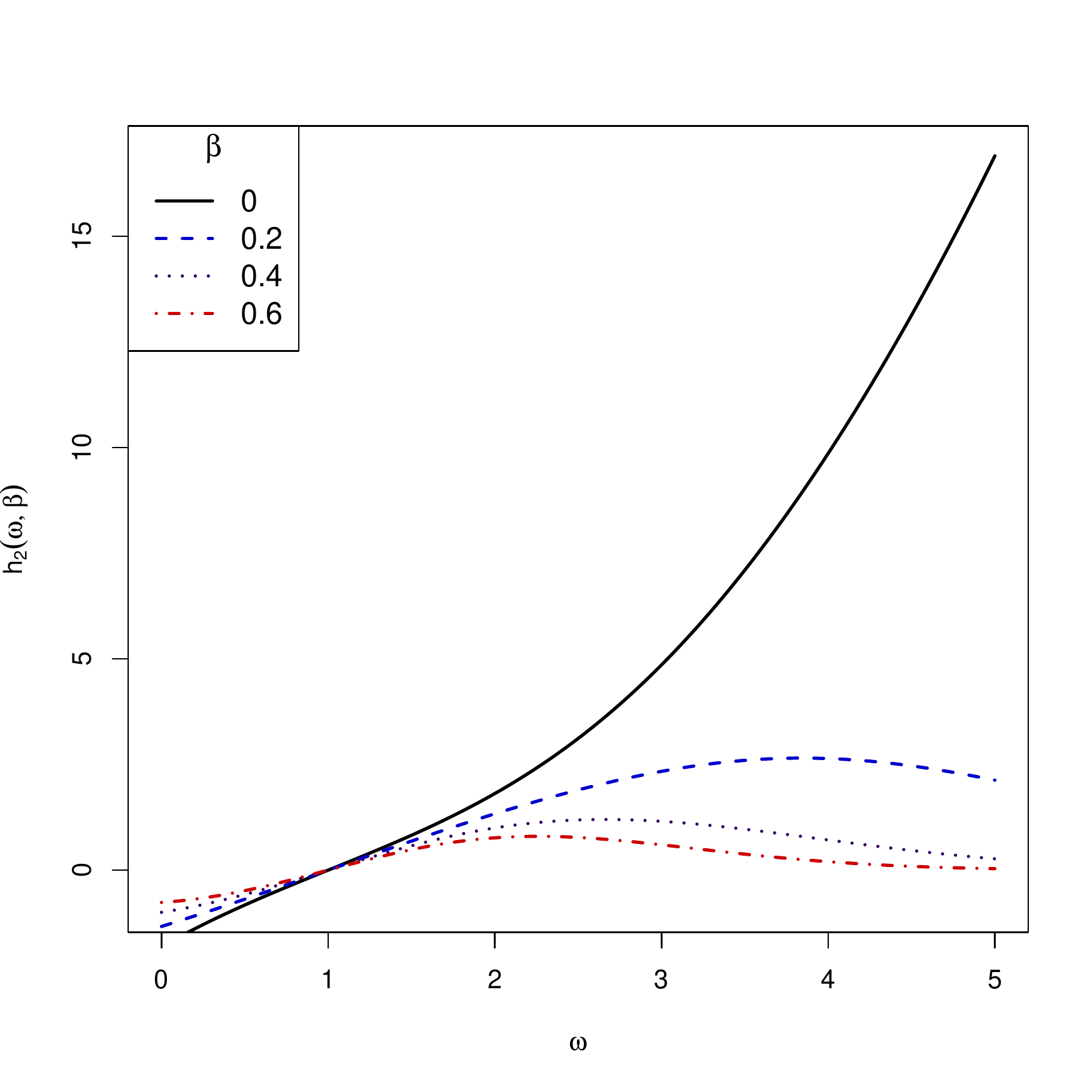}\\
	\includegraphics[scale=0.4]{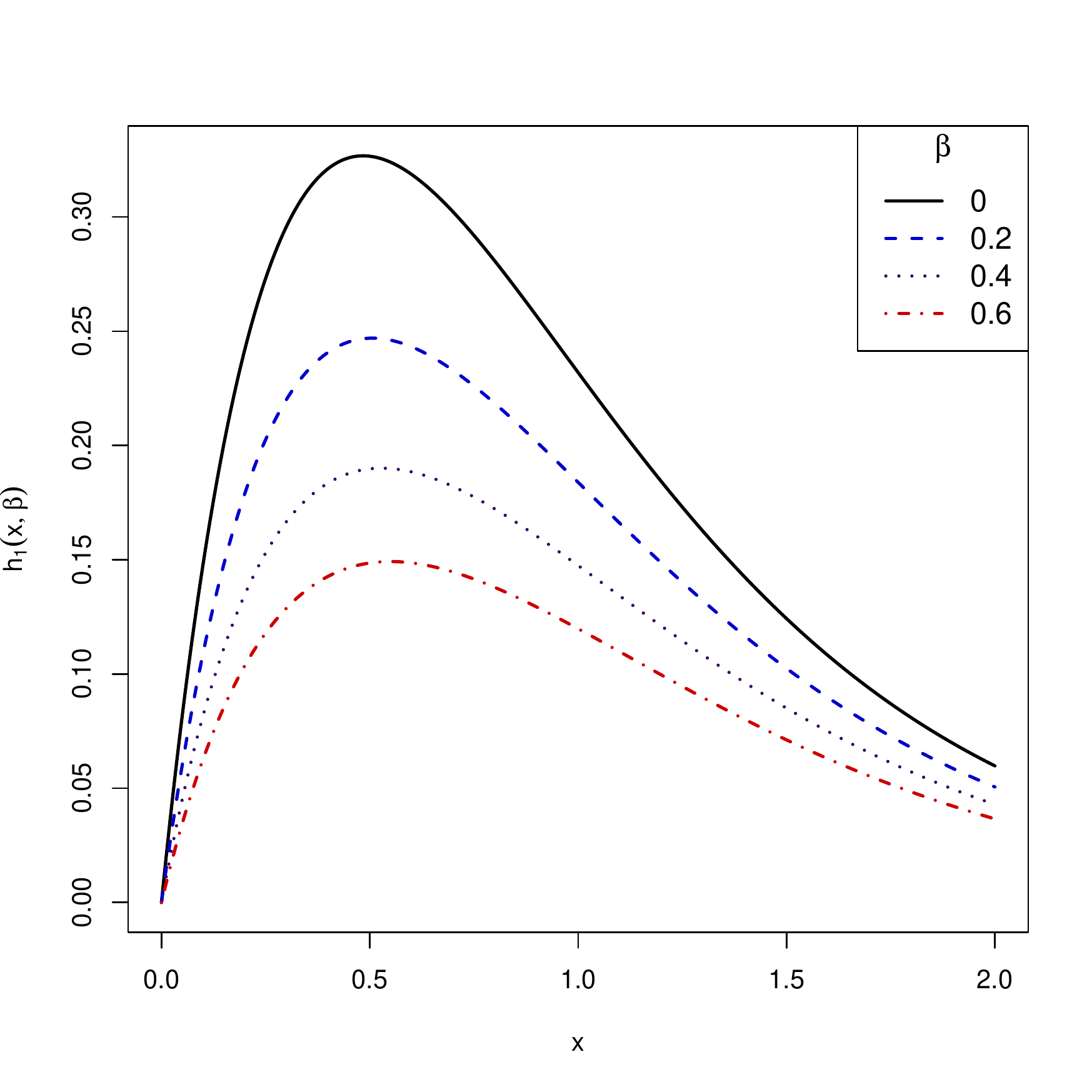}&
	\includegraphics[scale=0.4]{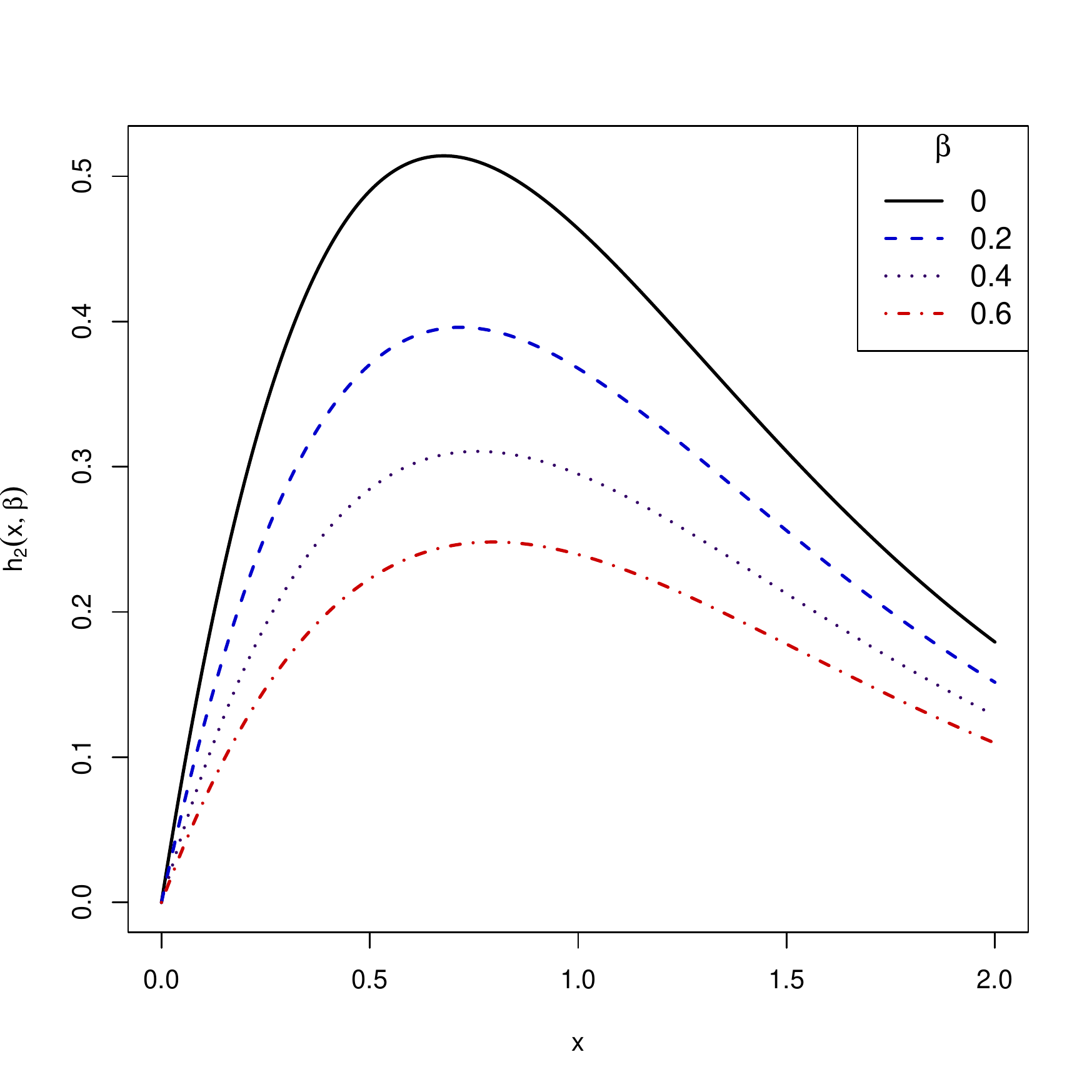}\\
	\includegraphics[scale=0.35]{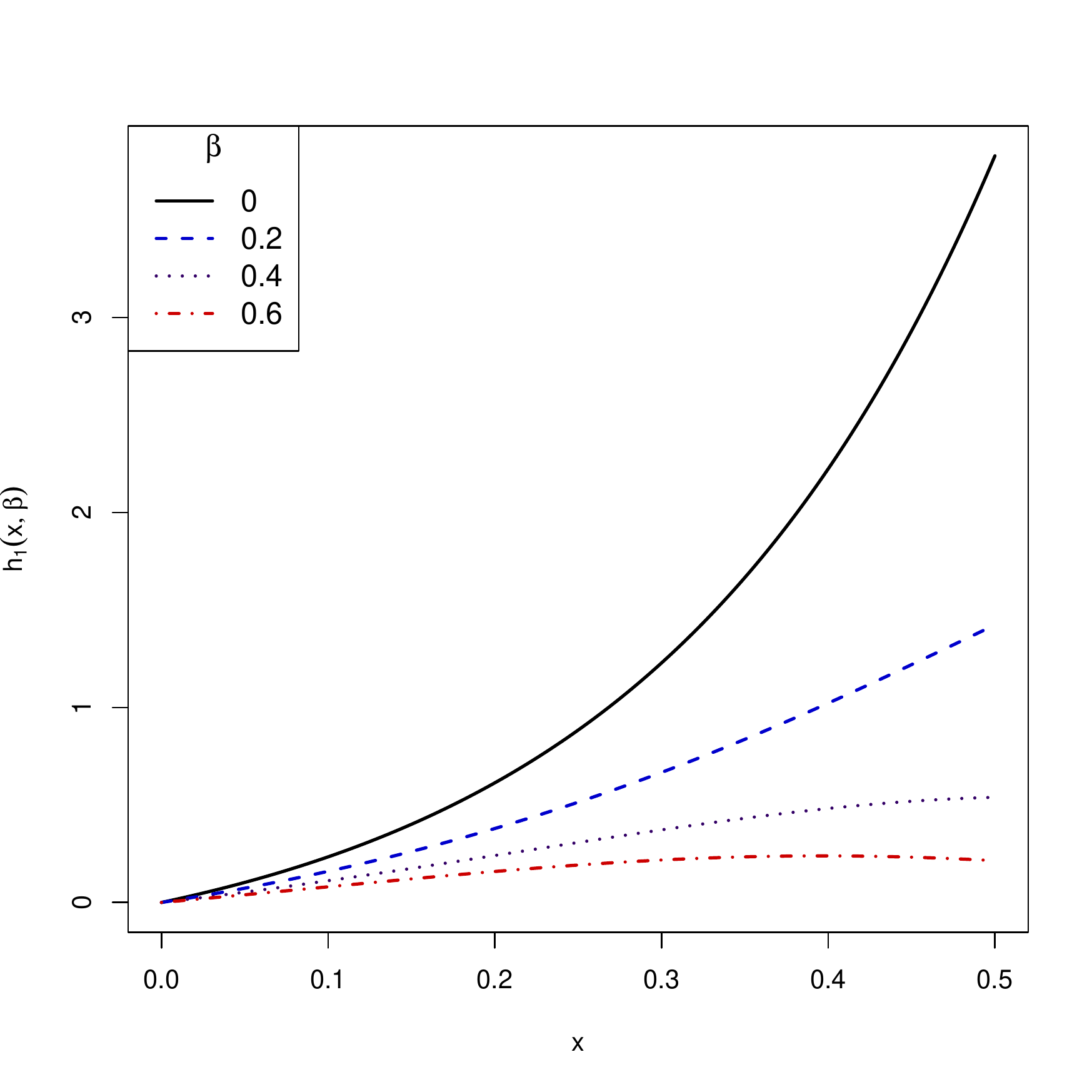}&
	\includegraphics[scale=0.35]{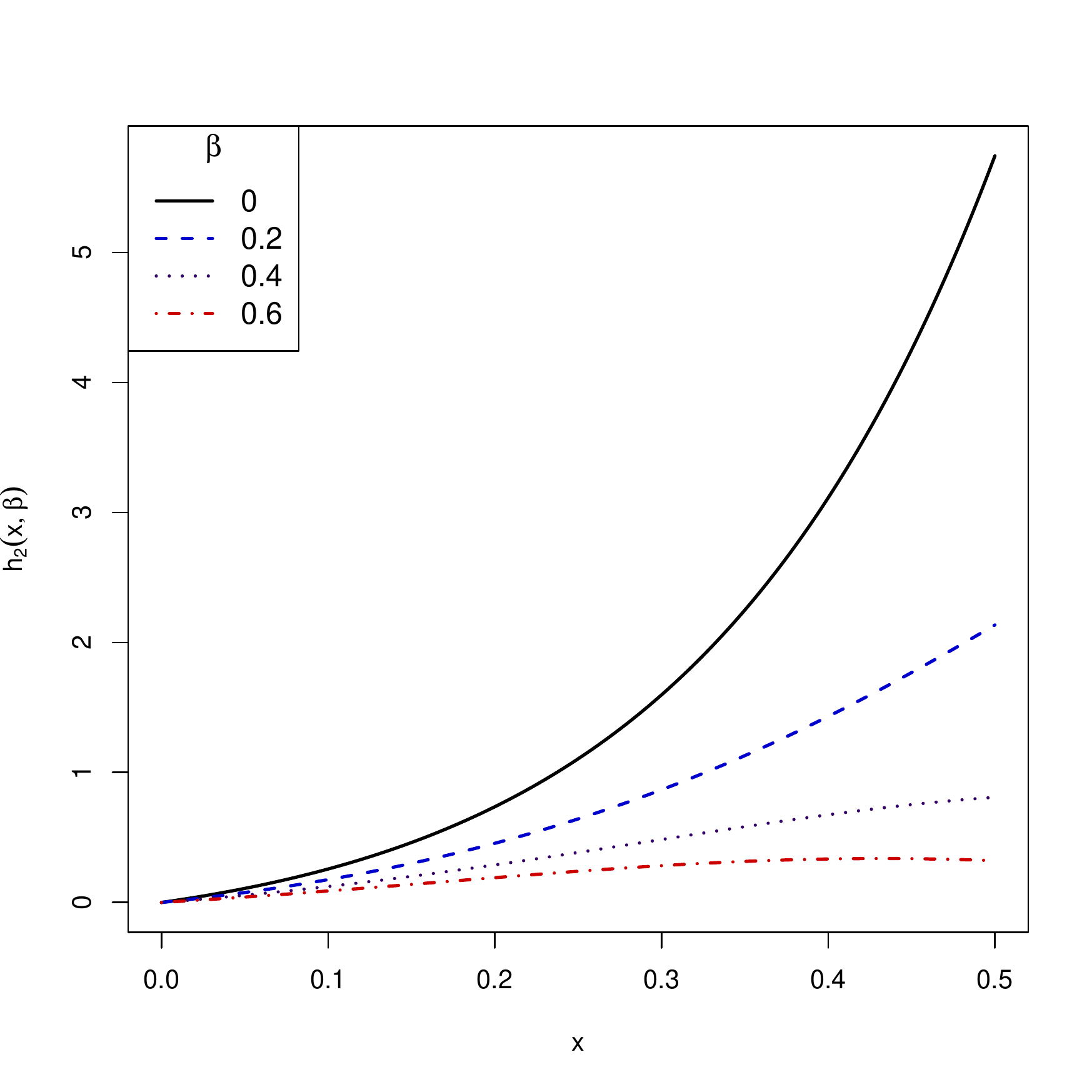}\\
	\end{tabular} 
	\caption{Effect of the change of $\omega$ and $x$ in $h_1(\omega,x,\boldsymbol{\theta},\beta)$ (left) and $h_2(\omega,x,\boldsymbol{\theta},\beta)$ (right) for different values of the tuning parameter $\beta$.}
	\label{fig:IF}
\end{figure}

%\begin{remark}
Let 
\begin{align}
h_1(\omega,x,\boldsymbol{\theta},\beta)&=\frac{1}{\sigma}\phi\left(\frac{\omega-\mu}{\sigma} \right)\left[\Phi\left(\frac{\omega-\mu}{\sigma} \right)^{\beta-1}+\left(1-\Phi\left(\frac{\omega-\mu}{\sigma} \right)\right)^{\beta-1} \right]x,\label{eq:i1}\\
h_2(\omega,x,\boldsymbol{\theta},\beta)&=\frac{\omega-\mu}{\sigma}\phi\left(\frac{\omega-\mu}{\sigma} \right)\left[\Phi\left(\frac{\omega-\mu}{\sigma} \right)^{\beta-1}+\left(1-\Phi\left(\frac{\omega-\mu}{\sigma} \right)\right)^{\beta-1} \right]x,\label{eq:i2}
\end{align}
be the factors of the influence function of $\boldsymbol{\theta}$ given in (\ref{eq:robust_IF_FINAL}) and (\ref{eq:robust_IF_FINAL_TOTAL}). Based on these, we may be comment on conditions for boundedness of the IFs presented here, either with respect to an observation or with respect to all the observations, that they are bounded on $t_{i_0,k}$ or $\boldsymbol{t}$; but, if \ $\beta =0$ the norm of the  influence functions can be very large on $(x,\omega)$, in comparison with $\beta >0$, as  can be seen in Figure \ref{fig:IF} for the case of only one factor of stress. 

Specifically to see the effect of $\omega$ in (\ref{eq:i1}) and (\ref{eq:i2}), we fix $\mu=1$, $\sigma=1$ and $x=1$. When $\omega$ increases, the classical MLE becomes clearly non-robust, which provides a clear justification for the need of ALT plans.  On the other hand, to see the effect of $x$ in (\ref{eq:i1}) and (\ref{eq:i2}), we take $\omega=1$ with $\boldsymbol{\theta}^T=(0,-1,0,1)$ or $\boldsymbol{\theta}^T=(0,-1,0,-1)$. Note that we are assuming $x$ to be positive, incrementing the stress as $x\rightarrow \infty$. Thus, we set $a_1<0$ to ensure an increment of the stress implies a reduction of the expected lifetime in (\ref{eq:E}).  Weighted minimum DPDs with $\beta>0$ present a more robust behaviour than the classical MLE.

%since
%
%\begin{align*}
%	\lim_{\substack{ x\rightarrow -\infty \\a_1>0 \\b_1>0}}|h_1(\omega,x,\boldsymbol{\theta},\beta)|&=\lim_{\substack{ x\rightarrow -\infty\\a_1>0 \\b_1>0} }|h_2(\omega,x,\boldsymbol{\theta},\beta)|=\lim_{\substack{ x\rightarrow \infty \\a_1>0 \\b_1<0}}|h_1(\omega,x,\boldsymbol{\theta},\beta)|=\lim_{\substack{ x\rightarrow \infty\\a_1>0 \\b_1<0} }|h_2(\omega,x,\boldsymbol{\theta},\beta)|\\
%		&=\lim_{\omega\rightarrow \infty }|h_1(\omega,x,\boldsymbol{\theta},\beta)|=\lim_{\omega\rightarrow \infty}|h_2(\omega,x,\boldsymbol{\theta},\beta)|
%		 \ \left\{ \begin{array}{cc}
%		= \infty , & \text{if }\beta =0 \\ 
%		<\infty , & \text{if }\beta >0%
%	\end{array}
%	\right. ,
%\end{align*}%

%  This implies that the proposed weighted minimum DPD estimators with $\beta>0$ are robust against large inspection times, but the classical MLE is clearly non-robust, which justifies the need of ALT plans. Same happens for leverage points under some particular conditions.
%\end{remark}

\subsection{Influence Function of the Wald-type tests}

The functional associated with the Wald-type statistics for testing the composite null hypothesis in (\ref{eq:nullhyp}) is given, ignoring the multiplier $K$, by
\begin{align*}
W_K(\boldsymbol{U}_{\beta})=(\boldsymbol{M}^T\boldsymbol{U}_{\beta}-\boldsymbol{m})^T\left[\boldsymbol{M}^T\boldsymbol{V}(\boldsymbol{U}_{\beta})\boldsymbol{M}\right]^{-1}(\boldsymbol{M}^T\boldsymbol{U}_{\beta}-\boldsymbol{m}),
\end{align*}
and the IF with respect to the $k-$th observation of the $i_0-$th group of observations, is then given by
\begin{align*}
IF(t_{i_0,k},W_K,F_{\boldsymbol{\theta}^0})=\left. \frac{\partial W_{K}(\boldsymbol{U}_{\beta}(F_{\boldsymbol{\theta}_{\varepsilon}^{i_0}}))}{\partial \varepsilon}\right|_{\varepsilon=0^+}=0.
\end{align*}
It, therefore, becomes necessary to consider the second-order IF, as presented in the following results.

\begin{result}
The second-order IF of the functional associated with the Wald-type test statistics, with respect to the $k-$th observation of the $i_0-$th group of observations, is given by

\begin{align*}
&IF_2(t_{i_0,k},W_K,F_{\boldsymbol{\theta}^0})=\left. \frac{\partial^2 W_{K}(\boldsymbol{U}_{\beta}(F_{\boldsymbol{\theta}_{\varepsilon}^{i_0}}))}{\partial \varepsilon^2}\right|_{\varepsilon=0^+}\\
&=2 \ IF(t_{i_0,k},\boldsymbol{U}_{\beta},F_{\boldsymbol{\theta}^0})(\boldsymbol{M}^T\boldsymbol{\theta}^0-\boldsymbol{m})^T\left[\boldsymbol{M}^T\boldsymbol{V}(\boldsymbol{\theta}^0)\boldsymbol{M}\right]^{-1}(\boldsymbol{M}^T\boldsymbol{\theta}^0-\boldsymbol{m})IF(t_{i_0,k},\boldsymbol{U}_{\beta},F_{\boldsymbol{\theta}^0}),
\end{align*}
where $IF(t_{i_0,k},\boldsymbol{U}_{\beta},F_{\boldsymbol{\theta}^0})$ is as given in (\ref{eq:robust_IF_FINAL}).
\end{result}

Similarly, for all the observations, we have the following result.

\begin{result}
The second-order IF of the functional associated with the Wald-type test statistics, with respect to all the observations, is given by

\begin{align*}
&IF_2(\boldsymbol{t},W_K,F_{\boldsymbol{\theta}^0})=\left. \frac{\partial^2 W_{K}(\boldsymbol{U}_{\beta}(F_{\boldsymbol{\theta}_{\varepsilon}}))}{\partial \varepsilon^2}\right|_{\varepsilon=0^+}\\
&=2 \ IF(\boldsymbol{t},\boldsymbol{U}_{\beta},F_{\boldsymbol{\theta}^0})(\boldsymbol{M}^T\boldsymbol{\theta}^0-\boldsymbol{m})^T\left[\boldsymbol{M}^T\boldsymbol{V}(\boldsymbol{\theta}^0)\boldsymbol{M}\right]^{-1}(\boldsymbol{M}^T\boldsymbol{\theta}^0-\boldsymbol{m})IF(\boldsymbol{t},\boldsymbol{U}_{\beta},F_{\boldsymbol{\theta}^0}),
\end{align*}
where $IF(\boldsymbol{t},\boldsymbol{U}_{\beta},F_{\boldsymbol{\theta}^0})$ is as given in (\ref{eq:robust_IF_FINAL_TOTAL}).
\end{result}

%\begin{remark}
Note that the second-order influence functions of the proposed Wald-type tests are quadratic functions of the corresponding IFs of the weighted minimum DPD estimator for any type of contamination. Therefore, the boundedness of the IFs of the weighted minimum DPD estimators at $\beta>0$ also indicates the boundedness of the IFs of the Wald-type test functional, implying their robustness against contamination.
%\end{remark}

%\clearpage
\section{Monte Carlo simulation study \label{sec:num}}
In this section,  a detailed simulation study is carried put to evaluate the performance of the proposed weighted minimum DPD estimators and the Wald-type tests under the assumption of lognormal lifetimes. We consider different scenarios with different sample sizes and under different degrees of contamination. The results are recorded and averaged over $S= 1,000$ simulation runs, using the R statistical software.

\subsection{Weighted minimum DPD estimators}
We evaluate the  robustness of the estimators by means of  standardized mean absolute error (SMAE)  of some specific parameters of interest for different values of the tuning parameter $\beta \in \{0,0.2,0.4,0.6 \}$.  After $S$ simulation runs,  the SMAE of a parameter of interest $\nu$ is computed as

$$
SMAE(\nu)=\frac{1}{S}\sum_{i=1}^S\left | \frac{\widehat{\nu}_i-\nu^0}{\nu^0} \right |,
$$
where $\nu^0$ is the true value of $\nu$ and $\widehat{\nu}_i$ is the estimated value of $\nu$ obtained from the $i$-th run.

%\subsubsection*{A. Balanced data}
The lifetimes of devices are simulated from the lognormal distribution, under three different stress conditions with one stress factor at three levels, taken to be $\{x_1,x_2,x_3\}=\{30,40,50\}$.  A balanced data with equal sample size for each group was considered. $K_i$ was taken to range from small to large sample sizes, and the model parameters were set to be 
$\boldsymbol{\theta} = (a_0 ,-0.1,-0.6,0.02)^T$, while $a_0$ was chosen to be $5.8$, $6.0$ and $6.2$ corresponding to devices with low, moderate, and high reliability, respectively. Then, all devices under each stress condition were tested at four different inspection times $\tau=\{\tau_1,\tau_2,\tau_3,\tau_4 \}$. To prevent many zero-observations in test groups, the inspection times were set as $\tau = \{5,10,15,20\}$ for the case of low reliability, $\tau= \{8,16,24,36\}$  for the case of moderate reliability, and $\tau = \{12,24,36,48\}$  for the case of high reliability.  Contaminated data were generated by setting $b_0$ as zero. SMAEs of the estimators are presented in Tables \ref{tb:low}, \ref{tb:moderate} and \ref{tb:high} for low, moderate and high reliability, respectively. As expected, contaminated data present higher errors than in the case of pure data. MLE presents the best behaviour in the case of pure data, while the minimum DPD estimators with $\beta>0$ outperform MLE in the presence of contamination.

\begin{figure*}[h]
\centering
\begin{tabular}{cc}
\includegraphics[scale=0.42]{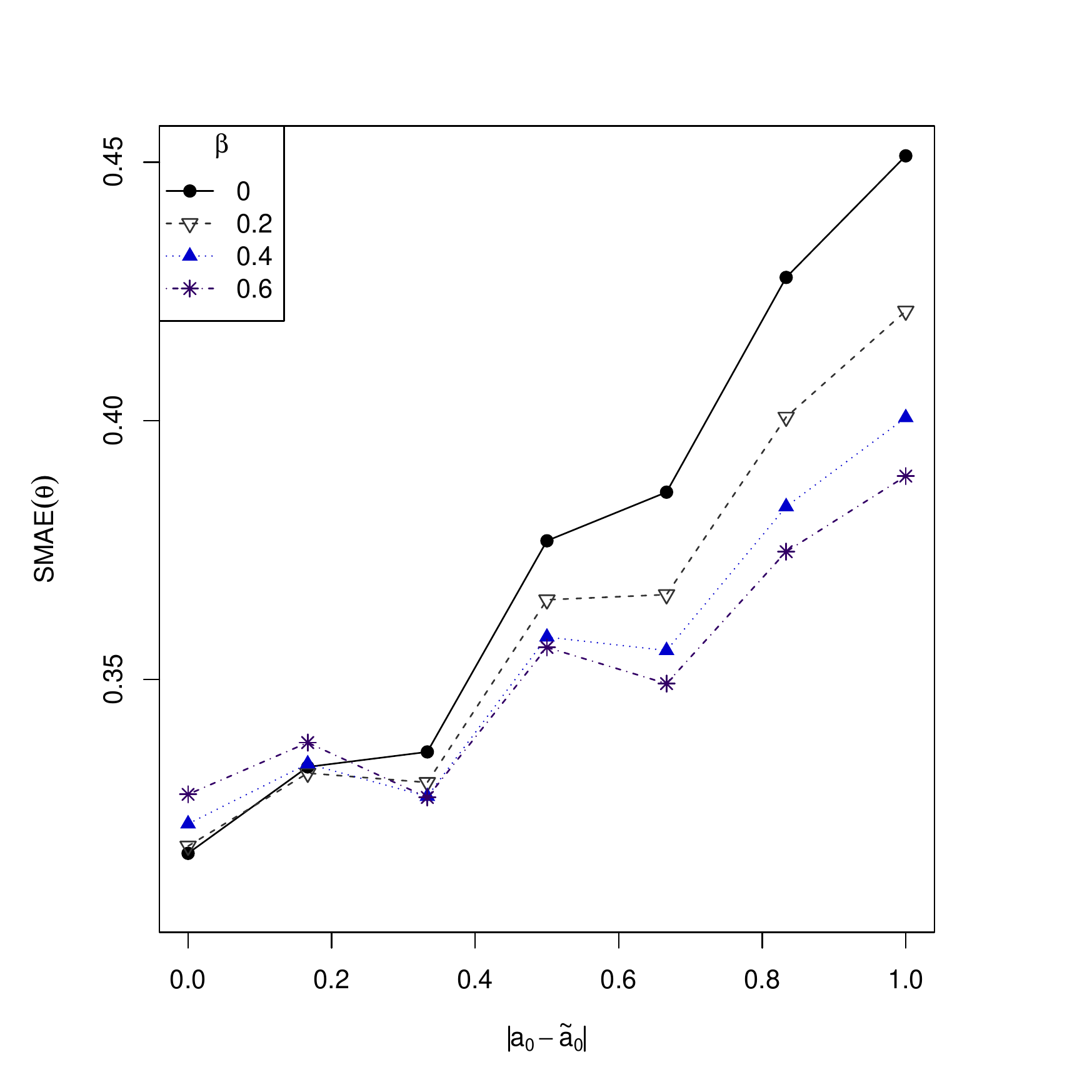}&
\includegraphics[scale=0.42]{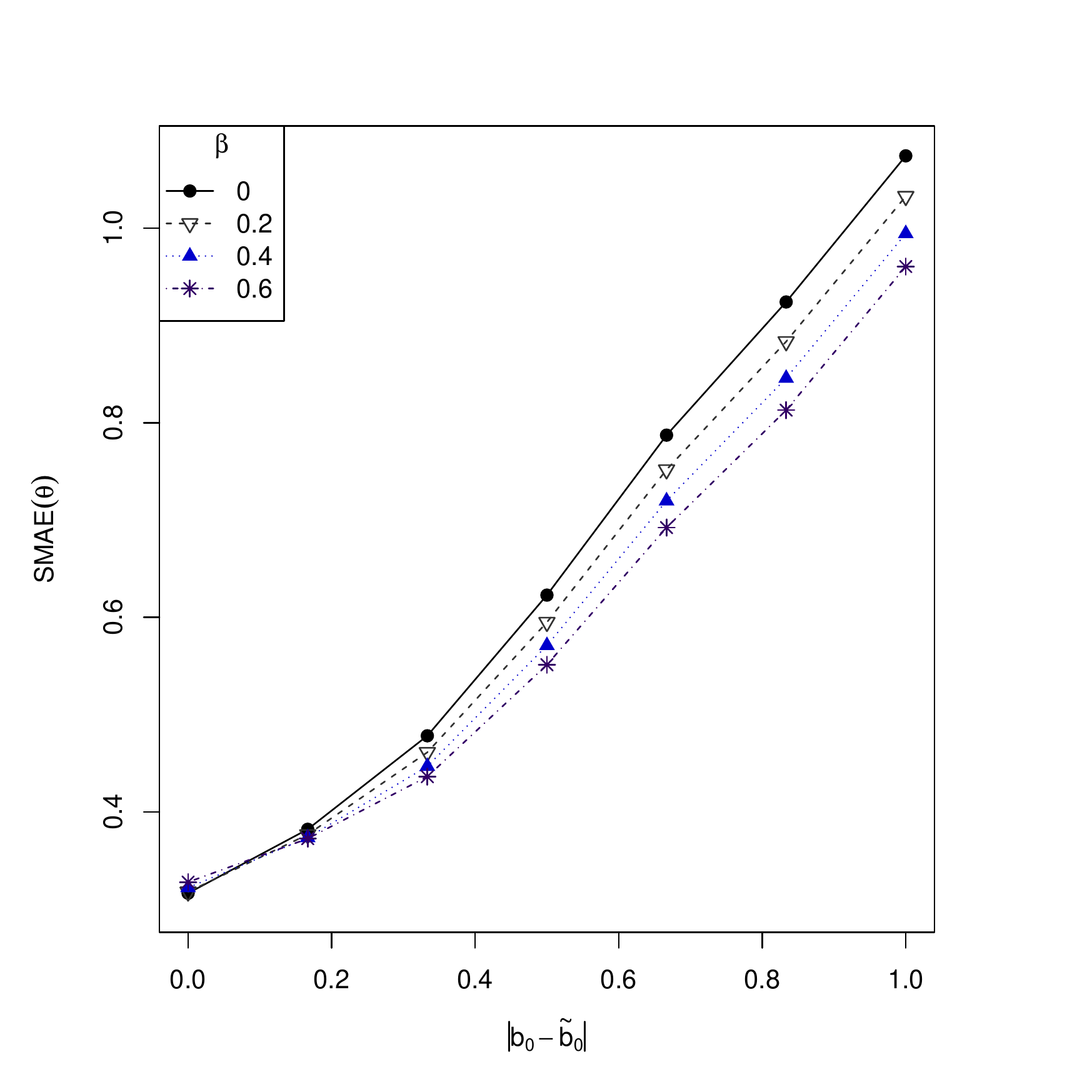}\\
\end{tabular} 
\caption{SMAEs of the parameter vector $\boldsymbol{\theta}$ for different levels of contamination. \label{fig:change}}
\end{figure*}

To better evaluate  the effect of contamination, we compute the SMAE of $\boldsymbol{\theta}$ for the moderate-reliability scenario defined previously with $K_i=80$ and under the contamination of parameters $a_0$ and $b_0$ chosen to be  $\tilde{a}_0\in [6,5]$ and $\tilde{b}_0\in [-0.6,0.4]$, as presented in Figure \ref{fig:change}.  As contamination increases, the precision of estimation decreases. Minimum DPD estimators with $\beta>0$ once again demostrate that they are more robust than the MLEs.
%\subsubsection*{B. Unbalanced data}
%In this scenario, an unbalanced data with different sample size in each condition is considered (see Table \ref{table:lognormal_ALTnb}). The data are generated with $\boldsymbol{\theta}=(6.0,-0.1,-0.6,0.02)^T$. To contamiante the data, the first cell is generated by $\tilde{\boldsymbol{\theta}}=(6.0,-0.1,-0.6,\tilde{b}_2)^T$, with $(1-\tilde{b}_2/b_2) \in (0,...,1)$. Note that when pure data are considered this value is equal to $0$. COMPORBAR QUE ESTA BIEN, SI ESO CAMBIAR OTRO PARAMETRO. Bias and RMSE of the parameter vector $\boldsymbol{\theta}$  are presented in top of Figure ???.
%
%
%\begin{table}[!!h!!!!]
%\caption{ALT plan, unbalanced data.\label{table:lognormal_ALTnb}}
%\center
%\begin{tabular}{c cccc}
%\hline
%i & $x_{i1}$  & $IT_i$ & $K_i$ \\ 
%\hline 
%1 & 30  & 12  & 60 \\ 
%2 & 40  & 12  & 40 \\ 
%3 & 50  & 12  & 20 \\ 
%4 & 30  & 16 & 60 \\ 
%5 & 40  & 16 & 20 \\ 
%6 & 50  & 16 & 20 \\ 
%7 & 30  & 24 & 40 \\ 
%8 & 40  & 24 & 20 \\ 
%9 & 50  & 24 & 20 \\ 
%\hline
%\end{tabular} 
%\end{table}

\subsection{Confidence intervals}

The coverage probability (CP) and average width (AW) are used for evaluating the $90\%$ confidence intervals of reliability and expected lifetime under normal operating conditions for different sample sizes  and the moderate-reliability scenario defined above. 

The results presented in Tables \ref{tb:conf1} and \ref{tb:conf2} indicate that the logit and arsech approaches outperform the  asymptotic confidence interval for reliability, for both pure and contaminated data. For the expected lifetime, there is not a significant difference between the asymptotic confidence interval and the log approach. But the latter is slightly more robust, but accompanied with an increase in the AW.

\subsection{Wald-type tests}

To evaluate the performance of the proposed Wald-type tests, we consider the scenario of moderate reliability mentioned in the last section. We consider the testing problem
\begin{equation}
H_0:a_0=6.0 \quad \text{against} \quad H_1:a_0\neq 6.0.
\end{equation}

\begin{figure*}[h]
\centering
\begin{tabular}{cc}
\includegraphics[scale=0.42]{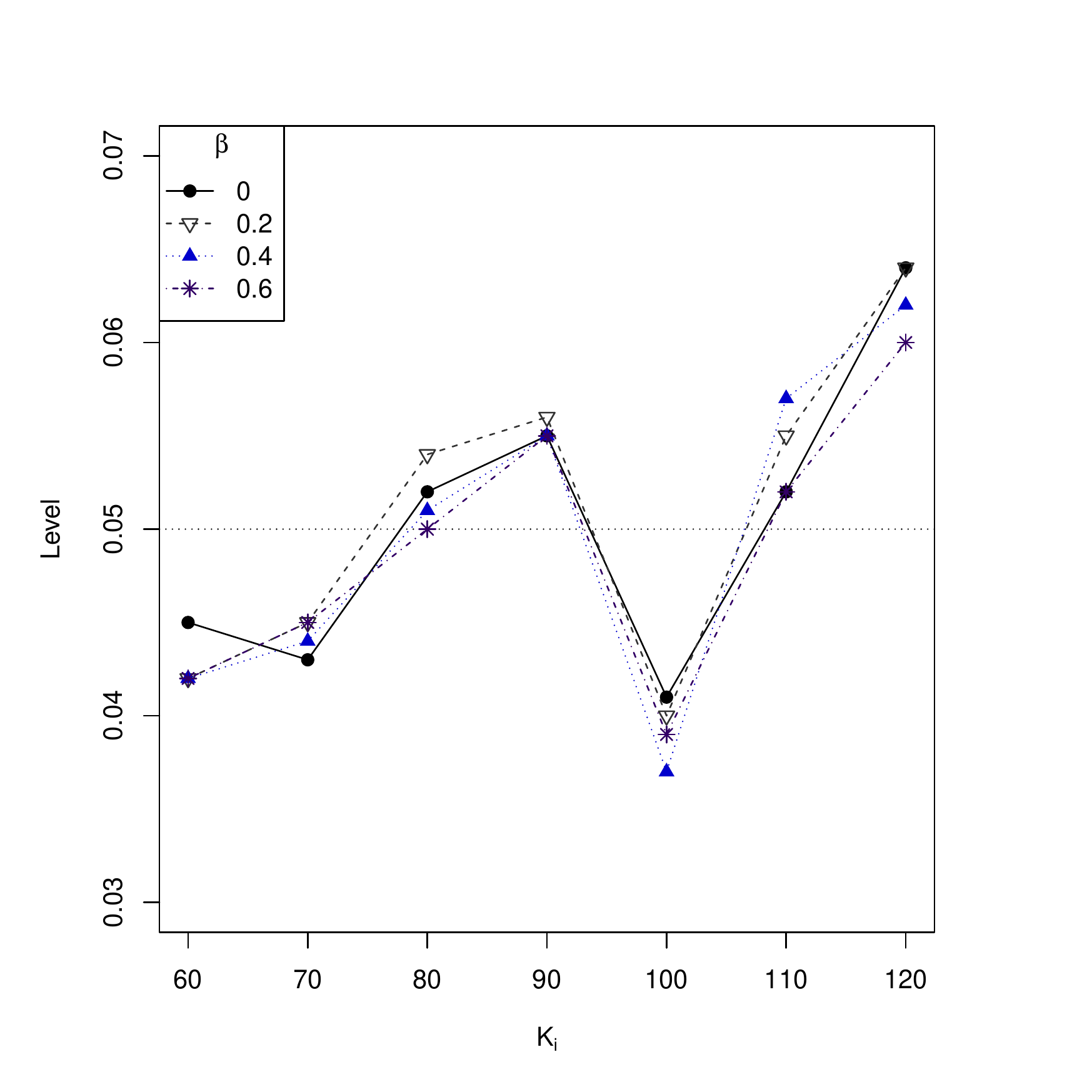}&
\includegraphics[scale=0.42]{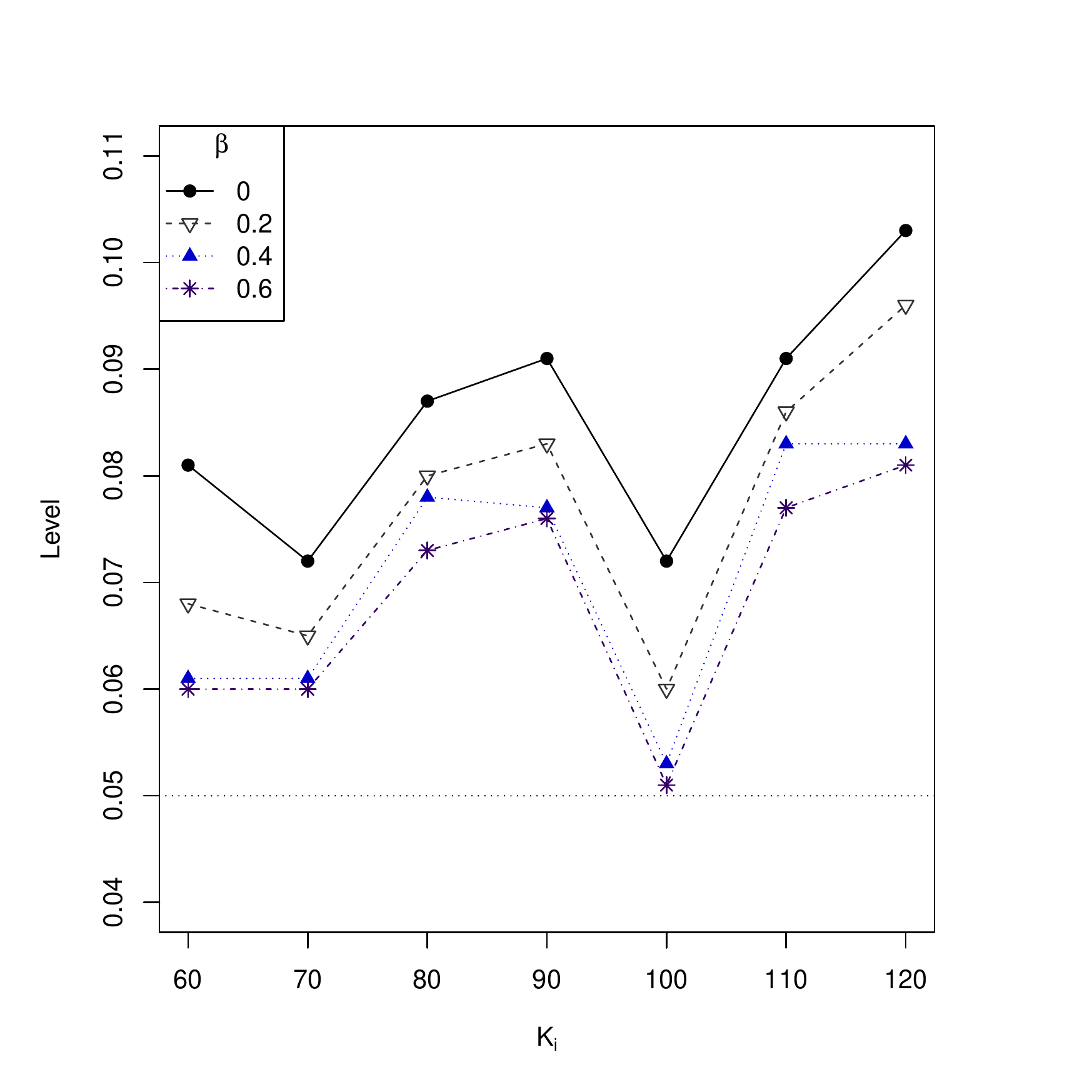}\\
\includegraphics[scale=0.42]{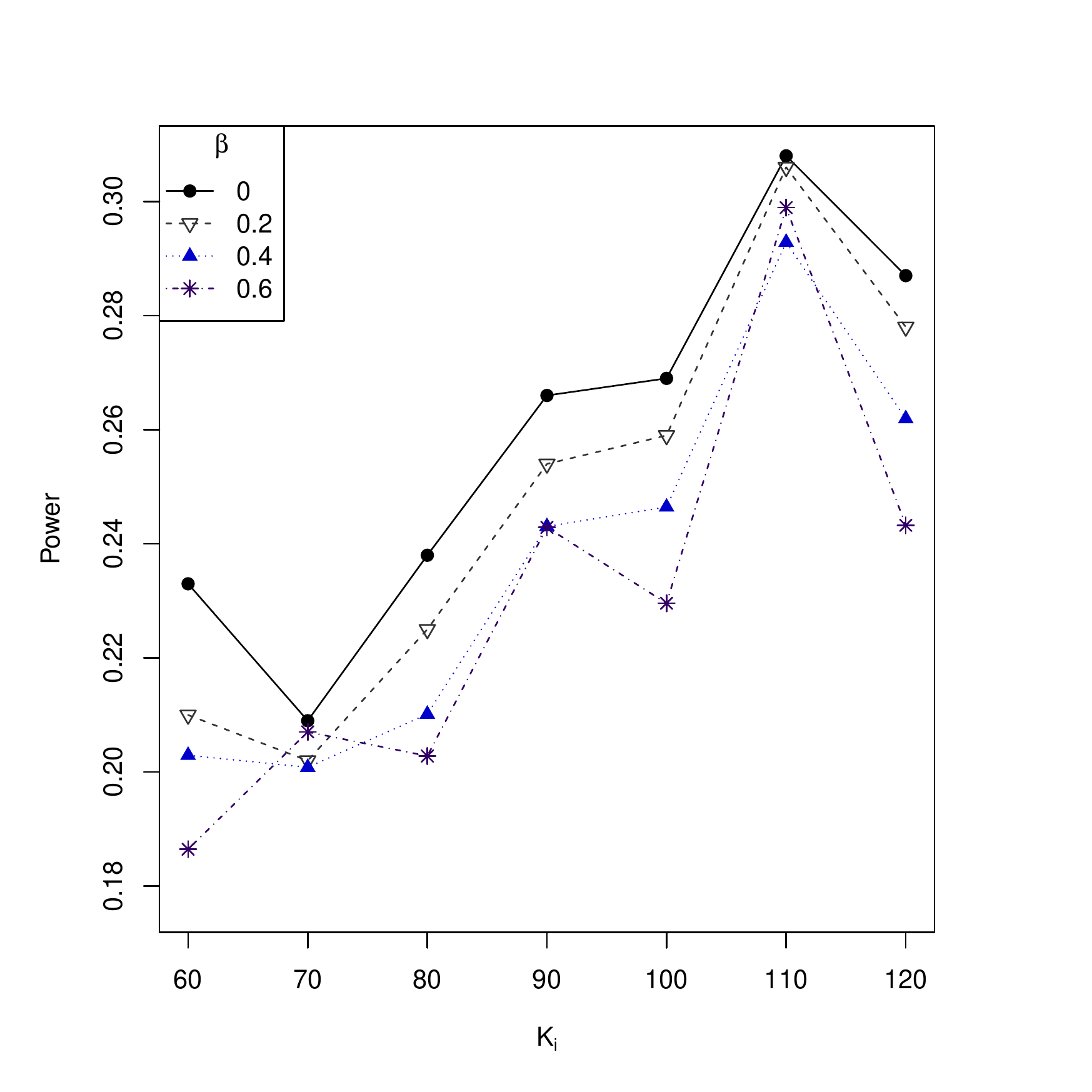}&
\includegraphics[scale=0.42]{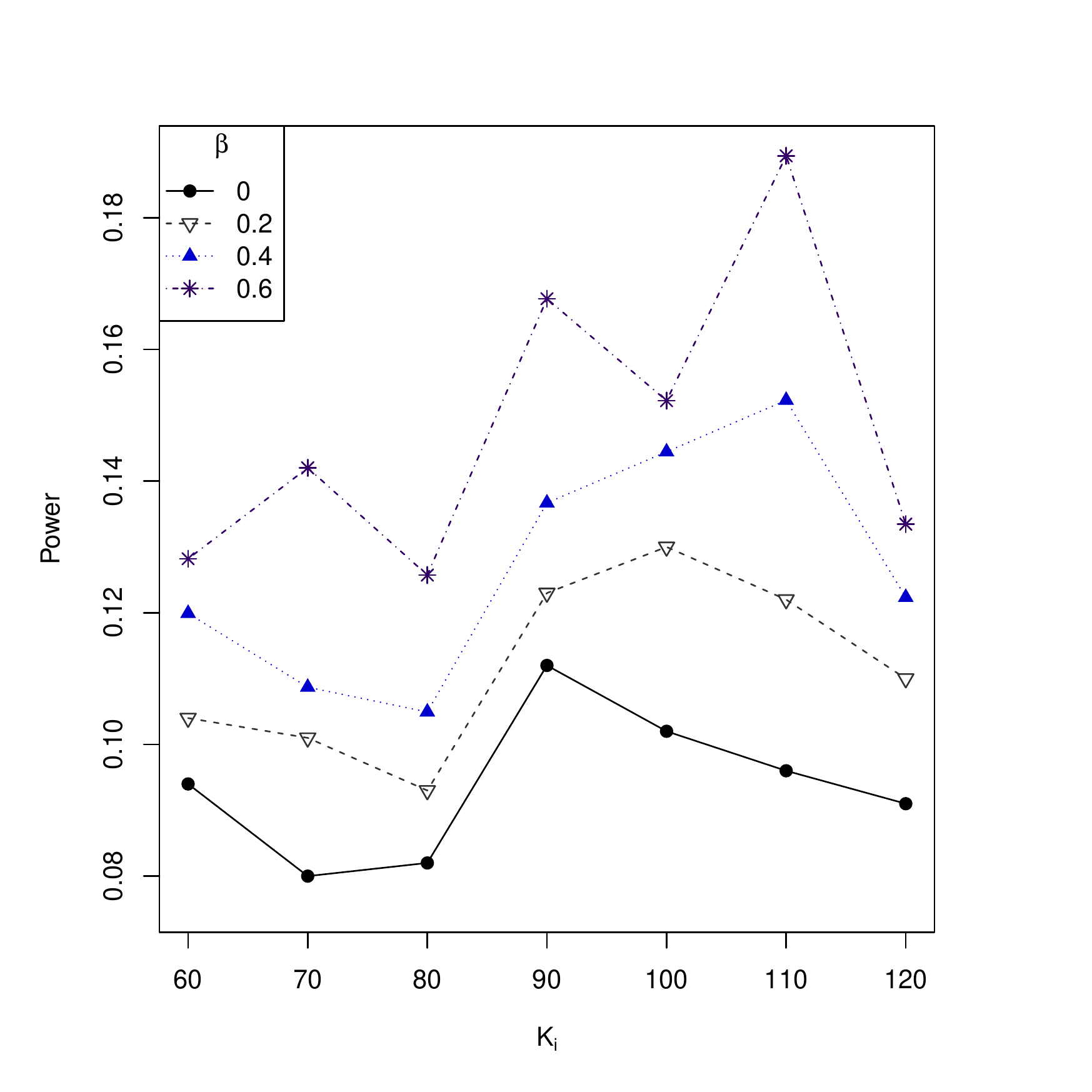}\\
\end{tabular} 
\caption{Empirical levels  and powers for pure (left) and contaminated (right) data. \label{fig:WALD}}
\end{figure*}

We first evaluate the empirical levels, measured as the proportion of test statistics exceeding the corresponding chi-square critical value for a nominal significant level of $\alpha=0.05$. The empirical powers are computed in a similar way, with $a_0^0=5$. The corresponding results are shown in  Figure \ref{fig:WALD} for both pure  and contaminated data (left and right, respectively).  To contaminate the data here, we set $\tilde{a}_0$ as $5.6$. For pure data, the whole family of Wald-type tests has their levels very close to the nominal level. However, when the data get contaminated, Wald-type tests  with $\beta>0$ have much more stable robustness properties. With respect to the power, classical Wald test outperforms other tests for pure data, but is less robust  when contamination is present.

%\subsection{Discussion of results}

%\subsection{Trade-off between robustness and efficiency}
%PODRIAMOS PONER AQUI LOS VALORES DE LOS ARE

\begin{table}[p]\renewcommand{\arraystretch}{1.1}\tabcolsep3.2pt
\caption{Mean Square Error of estimates of some parameters of interest at various levels of reliability and different sample sizes. Low reliability case. \label{tb:low}}
 \small
\center
\begin{tabular}{lr l rrrr l rrrr}
\hline
\multicolumn{2}{l}{Size} & &\multicolumn{4}{l}{Pure data} & &\multicolumn{4}{l}{Contaminated data}\\
\cline{1-2}  \cline{4-7} \cline{9-12}
	 & True value  && $\beta=0$ & $\beta=0.2$ & $\beta=0.4$ & $\beta=0.6$ && $\beta=0$ & $\beta=0.2$ & $\beta=0.4$ & $\beta=0.6$ \\ \hline
$K_i=50$ &       &  &         &     &     &          &        &        &             \\ \cline{1-2} 
$a_0$  & 5.8000  && 0.11043 & 0.10998 & 0.11121 & 0.11244 && 0.19983 & 0.18969 & 0.18078 & 0.17416 \\
$a_1$ & -0.1000  && 0.19862 & 0.19792 & 0.20004 & 0.20264 && 0.37594 & 0.35573 & 0.33816 & 0.32510 \\
$b_0$ & -0.6000  && 0.79721 & 0.79567 & 0.80597 & 0.82651 && 1.37877 & 1.31916 & 1.26482 & 1.22463 \\
$b_1$  & 0.0200  && 0.62687 & 0.62877 & 0.64152 & 0.66236 && 1.33147 & 1.26703 & 1.20589 & 1.15780 \\
$\boldsymbol{\theta}$ & -  && 0.43328 & 0.43308 & 0.43968 & 0.45099 && 0.82150 & 0.78290 & 0.74741 & 0.72042 \\
$R(60,15)$      & 0.6093   && 0.26439 & 0.26402 & 0.26730 & 0.26945 && 0.42266 & 0.40963 & 0.39723 & 0.38748 \\
$E(15)$      & 96.9704 && 0.79662 & 0.91662 & 1.21543 & 2.18463 && 1.13724 & 1.05631 & 0.99568 & 0.95358 \\\hline
$K_i=80$ &       &  &         &     &     &          &        &        &             \\ \cline{1-2} 
$a_0$  & 5.8000  && 0.08459 & 0.08545 & 0.08668 & 0.08685 && 0.17929 & 0.16786 & 0.15790 & 0.15054 \\
$a_1$ & -0.1000  && 0.15300 & 0.15438 & 0.15632 & 0.15649 && 0.33903 & 0.31687 & 0.29735 & 0.28283 \\
$b_0$ & -0.6000  && 0.60755 & 0.60786 & 0.61688 & 0.62660 && 1.31003 & 1.24521 & 1.18285 & 1.13126 \\
$b_1$  & 0.0200  && 0.48175 & 0.48409 & 0.49447 & 0.50574 && 1.29470 & 1.22338 & 1.15544 & 1.09756 \\
$\boldsymbol{\theta}$ & -  && 0.33172 & 0.33295 & 0.33859 & 0.34392 && 0.78076 & 0.73833 & 0.69838 & 0.66555 \\
$R(60,15)$     & 0.6093   && 0.21098 & 0.21319 & 0.21610 & 0.21604 && 0.42420 & 0.40421 & 0.38484 & 0.37033 \\
$E(15)$    & 96.9704 && 0.38567 & 0.38170 & 0.38343 & 0.38624 && 0.71432 & 0.66372 & 0.62268 & 0.59262 \\\hline
$K_i=100$ &       &  &         &     &     &          &        &        &             \\ \cline{1-2} 
$a_0$  & 5.8000  && 0.07502 & 0.07567 & 0.07741 & 0.07828 && 0.17073 & 0.15889 & 0.14892 & 0.14067 \\
$a_1$ & -0.1000  && 0.13612 & 0.13730 & 0.14014 & 0.14179 && 0.32353 & 0.30044 & 0.28084 & 0.26466 \\
$b_0$ & -0.6000  && 0.54224 & 0.54110 & 0.54658 & 0.55511 && 1.31806 & 1.24891 & 1.18423 & 1.12792 \\
$b_1$  & 0.0200  && 0.42211 & 0.42225 & 0.42877 & 0.43827 && 1.30147 & 1.22676 & 1.15638 & 1.09336 \\
$\boldsymbol{\theta}$ & -  && 0.29387 & 0.29408 & 0.29823 & 0.30336 && 0.77845 & 0.73375 & 0.69259 & 0.65665 \\
$R(60,15)$     & 0.6093   && 0.19164 & 0.19357 & 0.19749 & 0.19986 && 0.42210 & 0.39900 & 0.37876 & 0.36107 \\
$E(15)$     & 96.9704 && 0.31673 & 0.31898 & 0.32507 & 0.32770 && 0.64980 & 0.60181 & 0.56363 & 0.53396\\ \hline
\end{tabular}
\end{table}

\begin{table}[p]\renewcommand{\arraystretch}{1.1}\tabcolsep3.2pt
\caption{Mean Square Error of estimates of some parameters of interest at various levels of reliability and different sample sizes. Moderate reliability case. \label{tb:moderate}}
 \small
\center
\begin{tabular}{lr l rrrr l rrrr}
\hline
\multicolumn{2}{l}{Size} & &\multicolumn{4}{l}{Pure data} & &\multicolumn{4}{l}{Contaminated data}\\
\cline{1-2}  \cline{4-7} \cline{9-12}
	 & True value  && $\beta=0$ & $\beta=0.2$ & $\beta=0.4$ & $\beta=0.6$ && $\beta=0$ & $\beta=0.2$ & $\beta=0.4$ & $\beta=0.6$ \\ \hline
$K_i=30$ &       &  &         &     &     &          &        &        &             \\ \cline{1-2} 
$a_0$  & 6.0000  && 0.09570 & 0.09616 & 0.09672 & 0.09717 && 0.16123 & 0.15441 & 0.14855 & 0.14373\\
$a_1$ & -0.1000  && 0.17331 & 0.17414 & 0.17555 & 0.17656 && 0.30632 & 0.29226 & 0.28067 & 0.27124\\
$b_0$ & -0.6000  && 0.76588 & 0.76668 & 0.77483 & 0.78724 && 1.32374 & 1.27801 & 1.23946 & 1.20873 \\
$b_1$  & 0.0200  && 0.58694 & 0.58958 & 0.59992 & 0.61644 && 1.25706 & 1.20370 & 1.15763 & 1.12114\\
$\boldsymbol{\theta}$ & -  && 0.40546 & 0.40664 & 0.41176 & 0.41935 && 0.76209 & 0.73209 & 0.70658 & 0.68621 \\
$R(60,15)$ & 0.7080 &&0.19231 & 0.19293 & 0.19448 & 0.19666 && 0.27855 & 0.27215 & 0.26660 & 0.26149\\
$E(15)$ & 118.4399 && 0.63691 & 0.63599 & 0.64412 & 0.66485 && 0.84629 & 0.80038 & 0.76545 & 0.74235\\\hline
$K_i=50$ &       &  &         &     &     &          &        &        &             \\\cline{1-2} 
$a_0$  & 6.0000  && 0.07632 & 0.07677 & 0.07734 & 0.07738 && 0.13877 & 0.13171 & 0.12581 & 0.12081\\
$a_1$ & -0.1000  && 0.13831 & 0.13936 & 0.14037 & 0.14069 && 0.26610 & 0.25168 & 0.23945 & 0.22915\\
$b_0$ & -0.6000  && 0.61349 & 0.61479 & 0.62013 & 0.62832 && 1.24753 & 1.19550 & 1.14908 & 1.10774\\
$b_1$  & 0.0200  && 0.46786 & 0.47194 & 0.47933 & 0.48983 && 1.21401 & 1.15441 & 1.10035 & 1.05152\\
$\boldsymbol{\theta}$ & -  && 0.32400 & 0.32571 & 0.32929 & 0.33405 && 0.71660 & 0.68332 & 0.65367 & 0.62731 \\
$R(60,15)$ & 0.7080 && 0.15933 & 0.16043 & 0.16169 & 0.16261 && 0.27267 & 0.26246 & 0.25399 & 0.24583\\ 
$E(15)$ & 118.4399 && 0.40792 & 0.40842 & 0.41048 & 0.41170 && 0.58178 & 0.55206 & 0.52696 & 0.50842\\\hline
$K_i=100$ &       &  &         &     &     &          &        &        &             \\\cline{1-2} 
$a_0$  & 6.0000 && 0.06828 & 0.06872 & 0.06978 & 0.07059 && 0.13134 & 0.12399 & 0.11745 & 0.11234 \\
$a_1$ & -0.10000 && 0.12401 & 0.12482 & 0.12641 & 0.12804 && 0.25190 & 0.23680 & 0.22347 & 0.21287 \\
$b_0$ & -0.6000 && 0.52818 & 0.52705 & 0.53146 & 0.53672 && 1.25422 & 1.19977 & 1.14965 & 1.10668\\
$b_1$  & 0.02000 && 0.40643 & 0.40564 & 0.40949 & 0.41644 && 1.22467 & 1.16388 & 1.10610 & 1.05597\\
$\boldsymbol{\theta}$ & -  && 0.28172 & 0.28156 & 0.28428 & 0.28795 && 0.71553 & 0.68111 & 0.64917 & 0.62197  \\
$R(60,15)$ & 0.7080 && 0.14675 & 0.14749 & 0.14957 & 0.15149 && 0.27315 & 0.26246 & 0.25240 & 0.24388\\ 
$E(15)$ &  118.4399 && 0.32007 & 0.32095 & 0.32461 & 0.32704 && 0.49666 & 0.46851 & 0.44544 & 0.42782\\\hline
\end{tabular}
\end{table}

\begin{table}[p]\renewcommand{\arraystretch}{1.1}\tabcolsep3.2pt
\caption{Mean Square Error of estimates of some parameters of interest at various levels of reliability and different sample sizes. High reliability case. \label{tb:high}}
 \small
\center
\begin{tabular}{lr l rrrr l rrrr}
\hline
\multicolumn{2}{l}{Size} & &\multicolumn{4}{l}{Pure data} & &\multicolumn{4}{l}{Contaminated data}\\
\cline{1-2}  \cline{4-7} \cline{9-12}
	 & True value  && $\beta=0$ & $\beta=0.2$ & $\beta=0.4$ & $\beta=0.6$ && $\beta=0$ & $\beta=0.2$ & $\beta=0.4$ & $\beta=0.6$ \\ \hline
$K_i=30$ &       &  &         &     &     &          &        &        &             \\ \cline{1-2} 
$a_0$  & 6.2000  && 0.08996 & 0.09088 & 0.09138 & 0.09197 && 0.13200 & 0.12833 & 0.12478 & 0.12192\\
$a_1$ & -0.1000  &&  0.16250 & 0.16404 & 0.16522 & 0.16642 && 0.24958 & 0.24179 & 0.23460 & 0.22861\\
$b_0$ & -0.6000  &&  0.77303 & 0.77453 & 0.78027 & 0.78982 && 1.29934 & 1.26680 & 1.24278 & 1.22847 \\
$b_1$  & 0.0200  &&  0.58364 & 0.58674 & 0.59442 & 0.60514 && 1.19338 & 1.15206 & 1.11768 & 1.09119\\
$\boldsymbol{\theta}$ & -  && 0.40228 & 0.40405 & 0.40782 & 0.41334 && 0.71858 & 0.69725 & 0.67996 & 0.66755\\
$R(15;60)$ & 0.7932 && 0.14379 & 0.14452 & 0.14551 & 0.14666 && 0.17712 & 0.17399 & 0.17135 & 0.16947\\
$E(15)$ & 144.6628 && 0.86043 & 0.86502 & 0.87026 & 0.87939 && 105.452 & 0.94573 & 0.88077 & 0.84203\\\hline
$K_i=50$ &       &  &         &     &     &          &        &        &             \\\cline{1-2} 
$a_0$  & 6.2000  && 0.07261 & 0.07315 & 0.07337 & 0.07333 && 0.10841 & 0.10474 & 0.10181 & 0.09915\\
$a_1$ & -0.1000  && 0.12960 & 0.13039 & 0.13069 & 0.13090 && 0.20633 & 0.19840 & 0.19183 & 0.18622\\
$b_0$ & -0.6000  &&  0.63278 & 0.63262 & 0.63835 & 0.64322 && 1.21525 & 1.17574 & 1.14379 & 1.11819 \\
$b_1$  & 0.0200  &&  0.47713 & 0.47706 & 0.48087 & 0.48600 && 1.14733 & 1.10086 & 1.06085 & 1.02765\\
$\boldsymbol{\theta}$ & -  && 0.32803 & 0.32831 & 0.33082 & 0.33336 && 0.66933 & 0.64493 & 0.62457 & 0.60780 \\
$R(15;60)$ & 0.7932 && 0.12330 & 0.12381 & 0.12371 & 0.12351 && 0.17075 & 0.16613 & 0.16228 & 0.15947\\ 
$E(15)$ & 144.6628 && 0.44060 & 0.44289 & 0.44501 & 0.44693 && 0.48049 & 0.46618 & 0.45525 & 0.44650\\\hline
$K_i=100$ &       &  &         &     &     &          &        &        &             \\\cline{1-2} 
$a_0$  & 6.2000  && 0.06201 & 0.06305 & 0.06327 & 0.06349 && 0.10028 & 0.09600 & 0.09247 & 0.08929\\
$a_1$ & -0.1000  && 0.11093 & 0.11248 & 0.11301 & 0.11341 && 0.19164 & 0.18242 & 0.17491 & 0.16837\\
$b_0$ & -0.6000  && 0.54893 & 0.54975 & 0.55234 & 0.55988 && 1.17898 & 1.14104 & 1.10818 & 1.08244 \\
$b_1$  & 0.0200  &&  0.41805 & 0.41983 & 0.42345 & 0.42988 && 1.13684 & 1.08996 & 1.04886 & 1.01434\\
$\boldsymbol{\theta}$ & -  && 0.28498 & 0.28628 & 0.28802 & 0.29167 && 0.65193 & 0.62736 & 0.60610 & 0.58861 \\
$R(15;60)$ & 0.7932 &&  0.10550 & 0.10682 & 0.10730 & 0.10785 && 0.16873 & 0.16354 & 0.15927 & 0.15546\\ 
$E(15)$ & 144.6628 &&  0.34205 & 0.34669 & 0.34836 & 0.35068 && 0.40811 & 0.39425 & 0.38297 & 0.37325\\\hline
\end{tabular}
\end{table}

 \begin{table}[h]\renewcommand{\arraystretch}{1.1}
\caption{Coverage Probabilities  for  reliabilities and expected lifetime at  different sample sizes.  \label{tb:conf1}} \tabcolsep3.2pt
 \small
\center
\begin{tabular}{lrl l rrrr l rrrr}
\hline
\multicolumn{3}{l}{Size} & &\multicolumn{4}{l}{Pure data} & &\multicolumn{4}{l}{Contaminated data}\\
\cline{1-3}  \cline{5-8} \cline{10-13}
	 & True value  & Method && $\beta=0$ & $0.2$ & $0.4$ & $0.6$ && $\beta=0$ & $0.2$ & $0.4$ & $0.6$ \\  \hline
$K_i=50$ &       &  &         &     &     &          &        &        &             \\ \cline{1-2} 
R(15;60)  & 0.7080     & asy    && 0.787 & 0.794 & 0.790 & 0.785 && 0.539 & 0.566 & 0.582 & 0.595  \\
   				  && logit  && 0.891 & 0.896 & 0.892 & 0.888 &&  0.977 & 0.974 & 0.953 & 0.951 \\
			      && arsech && 0.862 & 0.858 & 0.848 & 0.844 && 0.630 & 0.656 & 0.678 & 0.690  \\
E(15) & 118.4399  & asy   && 0.881 & 0.881 & 0.884 & 0.880 && 0.817 & 0.829 & 0.840 & 0.848\\
   			&    & log  && 0.828 & 0.823 & 0.825 & 0.828 && 0.843 & 0.836 & 0.834 & 0.833 \\
\hline 
$K_i=80$ &       &  &         &     &     &          &        &        &             \\ \cline{1-2} 
R(15;60)  & 0.7080     & asy    &&  0.817 & 0.820 & 0.820 & 0.816 && 0.497 & 0.527 & 0.574 & 0.592 \\
   				  && logit  && 0.901 & 0.903 & 0.902 & 0.903 && 0.996 & 0.991 & 0.987 & 0.984 \\
			      && arsech && 0.873 & 0.866 & 0.866 & 0.868 && 0.624 & 0.658 & 0.681 & 0.695 \\
E(15) & 118.4399  & asy   && 0.889 & 0.886 & 0.884 & 0.887 && 0.821 & 0.831 & 0.833 & 0.838 \\
   			&    & log  && 0.857 & 0.855 & 0.855 & 0.853 && 0.878 & 0.878 & 0.872 & 0.865 \\ 	
 \hline 
$K_i=100$ &       &  &         &     &     &          &        &        &             \\ \cline{1-2}  			
R(15;60)  & 0.7080     & asy    && 0.836 & 0.836 & 0.833 & 0.828 && 0.478 & 0.518 & 0.545 & 0.574 \\
   				  && logit  && 0.911 & 0.916 & 0.913 & 0.910 && 0.999 & 0.996 & 0.996 & 0.995 \\
			      && arsech && 0.878 & 0.875 & 0.873 & 0.869 && 0.574 & 0.606 & 0.633 & 0.667 \\
E(15) & 118.4399  & asy   && 0.895 & 0.895 & 0.893 & 0.892 && 0.814 & 0.825 & 0.837 & 0.840 \\
   			&    & log  && 0.861 & 0.861 & 0.861 & 0.859 && 0.890 & 0.889 & 0.881 & 0.877 \\ 	
 \hline 
\end{tabular}
\end{table}

%%%%%%%%%%%

 \begin{table}[h]\renewcommand{\arraystretch}{1.1}
\caption{Average Widths of confidence intervals for  reliabilities and expected lifetime at  different sample sizes.  \label{tb:conf2}} \tabcolsep2.5pt
 \small
\center
\begin{tabular}{lrl l rrrr l rrrr}
\hline
\multicolumn{3}{l}{Size} & &\multicolumn{4}{l}{Pure data} & &\multicolumn{4}{l}{Contaminated data}\\
\cline{1-3}  \cline{5-8} \cline{10-13}
	 & True value  & Method && $\beta=0$ & $0.2$ & $0.4$ & $0.6$ && $\beta=0$ & $0.2$ & $0.4$ & $0.6$ \\  \hline
$K_i=50$ &       &  &         &     &     &          &        &        &             \\ \cline{1-2} 
R(15;60)  & 0.7080 & asy    && 0.362   & 0.362   & 0.364   & 0.365   && 0.325   & 0.334   & 0.341   & 0.347   \\
&& logit  && 0.378   & 0.380   & 0.383   & 0.386   && 0.545   & 0.524   & 0.511   & 0.502   \\
&& arsech && 0.310   & 0.310   & 0.310   & 0.310   && 0.233   & 0.242   & 0.249   & 0.255   \\
E(15) & 118.4399  & asy   && 148.30 & 148.33 & 149.15 & 150.60 && 161.66 & 159.13 & 157.19 & 155.956 \\
   			&    & log  && 162.09 & 162.06 & 163.01 & 164.71 && 181.40 & 177.53 & 174.64 & 172.707 \\
\hline 
$K_i=80$ &       &  &         &     &     &          &        &        &             \\ \cline{1-2} 
R(15;60)  & 0.7080 & asy  &&0.304 & 0.305 & 0.307 & 0.308 && 0.270 & 0.280 & 0.288   & 0.295 \\
&& logit  && 0.310   & 0.312   & 0.314   & 0.316   && 0.452   & 0.444   & 0.438   & 0.432  \\
&& arsech && 0.275   & 0.276   & 0.277   & 0.278   && 0.216   & 0.226   & 0.234   & 0.241 \\
E(15) & 118.4399  & asy   && 107.45 & 107.54 & 107.90 & 108.56 && 120.73 & 118.82 & 117.21 & 116.020 \\
   			&    & log  && 113.60 & 113.70 & 114.13 & 114.88 && 129.99 & 127.50 & 125.47 & 123.95 \\ 	
 \hline 
$K_i=100$ &       &  &         &     &     &          &        &        &             \\ \cline{1-2}  			
R(15;60)  & 0.7080 & asy   && 0.280   & 0.281   & 0.282   & 0.284   && 0.246   & 0.256   & 0.265   & 0.272 \\
&& logit  && 0.282   & 0.283   & 0.285   & 0.287   && 0.398   & 0.392   & 0.387   & 0.383 \\
&& arsech &&  0.258   & 0.259   & 0.260   & 0.261   && 0.205   & 0.214   & 0.223   & 0.229 \\
E(15) & 118.4399  & asy   && 90.93  & 90.87  & 91.264  & 91.68  && 105.06 & 103.29 & 101.79 & 100.619  \\
&& log  && 95.06  & 95.01  & 95.43  & 95.89  && 111.62 & 109.50 & 107.67 & 106.27 \\ 	
 \hline 
\end{tabular}
\end{table}

%\clearpage
\section{Illustrative examples \label{sec:ex}}

In this section, we present two real-life examples to illustrate all the methods of inference developed in the preceding sections: \\

%\subsection{Electro-explosive devices}
\noindent \textbf{Electro-explosive device data:} In this one-shot device testing experiment described in Balakrishnan and Ling (2012) (Table 1), the number of failures of $90$ electro-explosive devices tested at temperatures $x_i \in \{35,45,55 \}$ and inspection times $\tau_i \in\{10,20,30\}$ were collected. In this example, we have only one stress factor, with $I=9$ and $K_i=10$, for $i=1,\dots,I$.\\ 
%\subsection{Electric current data}
\noindent \textbf{Electric current data:} These data, presented in Balakrishnan and Ling (2012) (Table 2), present the number of failures on $120$ devices  placed under 12 different conditions in the experiment:  four accelerated conditions with higher-than-normal temperature $x_{i1} \in \{55,85\}$ and electric current $x_{i2} \in \{70,100\}$, and  three different inspection times $\tau_i\in \{2,5,8\}$. In this case, we have $J=2$, $I=12$ and $K_i=10$, for $i=1,\dots,I$.

\begin{figure*}[h]
\centering
\begin{tabular}{c}
\includegraphics[scale=0.5]{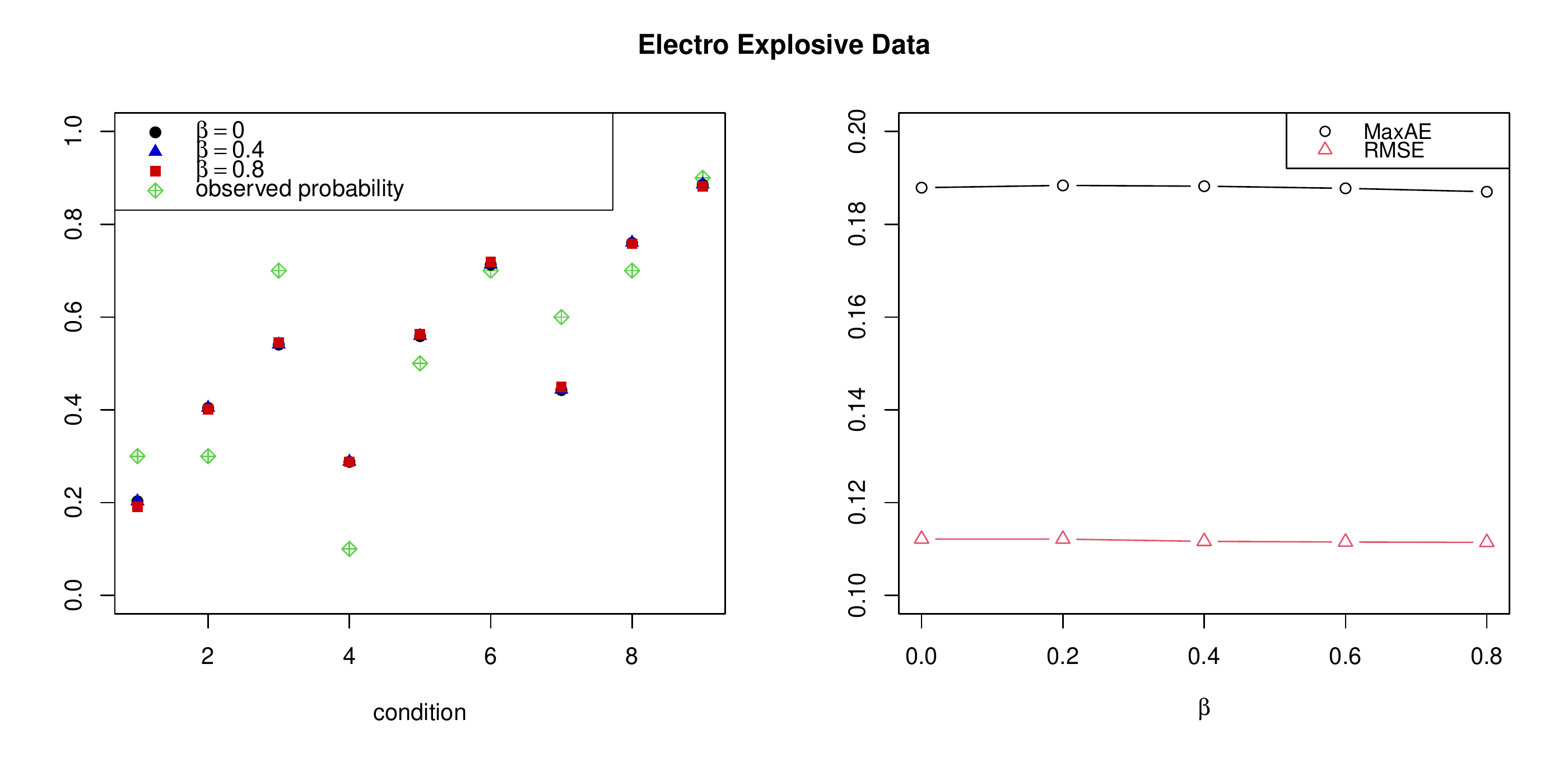}\\
\includegraphics[scale=0.5]{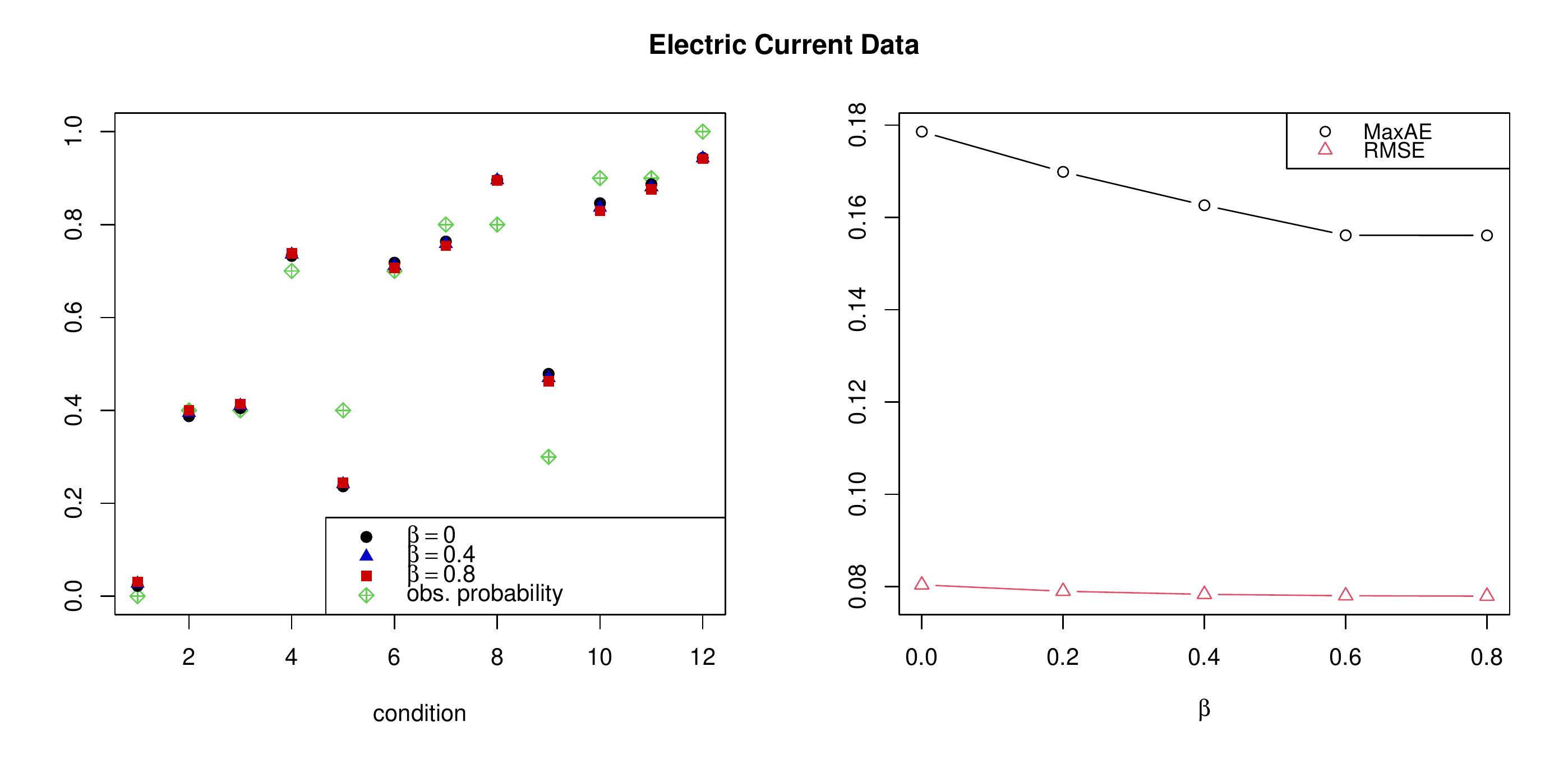}
\end{tabular} 
\caption{Electro-explosives  and Electric current examples. Left: estimated versus observed probabilities. Right: RMSEs and MaxAEs of the estimated probabilities. \label{fig:examples}}
\end{figure*}

Weighted minimum DPD estimators  were computed for different values of the tuning parameter $\beta$ and the estimated values of some specific parameters of interest are presented in Table \ref{ex:electro} (Electro-explosive data) and Table \ref{ex:electric} (Electric current data). For the CI of reliability, we use the logit-approach, while for the CI of mean lifetime, we use the log-approach. For both examples, the predicted probabilities are compared to the observed ones (left of Figure \ref{fig:examples}). 

In the preceding discussion, we noted that the robustness feature of the proposed weighted minimum DPD estimators usually increases with increasing $\beta$; but the  efficiency  decreases slightly. It seems, therefore, that a moderately large value of $\beta$ would provide the best trade-off for possibly contaminated data. However, a data-driven choice of $\beta$ may be more helpful in practice. In the context of one-shot device testing, Castilla and Chocano (2022) studied different methods for the  the choice of the ``optimal'' tuning parameter. The Warwick and Jones procedure (Warwick and Jones, 2005) and its iterative form (Basak et al., 2020) propose a minimization of the asymptotic mean square error (MSE) through the computation of the estimated asymptotic variance-covariance matrix of the model parameters in a grid of tuning parameters. However, as seen in    Castilla and Chocano (2022), minimizing a loss function which relates empirical and theoretical probabilities may provide easy-to-compute criteria. Particularly, the minimization of the maximum of the absolute errors (MaxAE) is seen to work well. We compute the MaxAE and RMSE for our examples, and we observe how for the Electric current data, the estimation with higher tuning parameters clearly reduces the MaxAE.

%Given a data set, the choice of the ``optimal'' tuning parameter is an important question for which several work h
%\clearpage

\begin{table}[]
\caption{Inferential results for electro-explosive device data  \label{ex:electro}}
\center
\begin{tabular}{lrrrrr}
\hline
Parameter & Estimate & Confidence Interval && Estimate & Confidence Interval\\
\hline
   & \multicolumn{2}{c}{$\beta=0$}& &\multicolumn{2}{c}{$\beta=0.2$}\\
\cline{2-3}  \cline{5-6} 
$a_0$  & 4.78801   & (3.21825, 6.35778)    &  & 4.78801   & (3.21861, 6.35742)    \\
$a_1$  & -0.04305  & (-0.07661, -0.00949)  &  & -0.04308  & (-0.07665, -0.00952)   \\
$b_0$  & 0.80430    & (-1.93904, 3.54764)    &  & 0.8043    & (-1.94143, 3.55002)    \\
$b_1$  & -0.01833  & (-0.07575, 0.03910)     &  & -0.01833  & (-0.07583, 0.03917)    \\
$R(25;10)$  & 0.84059   & (0.34212, 0.98164)    && 0.84049   & (0.34145, 0.98167)    \\
$R(35;10)$  & 0.82270    & (0.46616, 0.96102)    &  & 0.82254   & (0.46560, 0.96103)    \\
$E(25)$ & 111.16472 & (5.27000, 2344.89266) && 111.02344 & (5.27234, 2337.89866) \\  \hline
   & \multicolumn{2}{c}{$\beta=0.4$}& &\multicolumn{2}{c}{$\beta=0.6$}\\
\cline{2-3}  \cline{5-6} 
$a_0$ & 4.78858   & (3.25352, 6.32364)    &  & 4.78693   & (3.27006, 6.30380)     \\
$a_1$  & -0.04325  & (-0.07645, -0.01006)   &  & -0.04329  & (-0.07628, -0.01031)   \\
$b_0$ & 0.60442   & (-2.08826, 3.29709)    &  & 0.50556   & (-2.16925, 3.18036)    \\
$b_1$ & -0.01420   & (-0.07111, 0.04270)     &  & -0.0122   & (-0.06901, 0.04461)    \\
$R(25;10)$& 0.86317   & (0.32925, 0.98782)    &  & 0.87435   & (0.32045, 0.99036)    \\
$R(35;10)$ & 0.83996   & (0.46974, 0.96884)    &  & 0.84876   & (0.47042, 0.97257)   \\
$E(25)$  & 92.80086  & (7.57122, 1137.46539) &  & 85.75138  & (8.77878, 837.62209) \\ \hline 
\end{tabular}
\end{table}

\begin{table}[]
\caption{Inferential results for electric current data \label{ex:electric}}
\center
\begin{tabular}{lrrrrr}
\hline
Parameter & Estimate & Confidence Interval && Estimate & Confidence Interval\\
\hline
   & \multicolumn{2}{c}{$\beta=0$}& &\multicolumn{2}{c}{$\beta=0.2$}\\
\cline{2-3}  \cline{5-6} 
$a_0$&6.91992    & (4.61124,   9.22859)    && 7.04428    & (4.59463,    9.49392)    \\
$a_1$&-0.03979   & (-0.05979, -0.01980)   && -0.04062   & (-0.06159,   -0.01965)   \\
$a_2$&-0.03734   & (-0.05785,  -0.01683)   && -0.03820   & (-0.05989,   -0.01651)   \\
$b_0$&-1.84593   & (-4.00264,  0.31079)    && -1.71032   & (-3.89667,   0.47603)    \\
$b_1$&0.01003    & (-0.01212,  0.03219)    && 0.00914    & (-0.01358,   0.03187)    \\
$b_2$&0.01354    & (-0.00837,  0.03546)    && 0.01314    & (-0.00936,   0.03563)    \\
$R(25,35,60)$&0.94600    & (0.00124 ,  100.000)    && 0.95153    & (0.00153,    100.000)    \\
$R(35,50,60)$&0.46376    & (0.01032,   0.98625)    && 0.52216    & (0.01568,    0.98684)    \\
$E(25,35)$&106.83129 & (27.56069, 414.10155) && 116.33824 & (27.38673,  494.20227) \\ \hline
   & \multicolumn{2}{c}{$\beta=0.4$}& &\multicolumn{2}{c}{$\beta=0.6$}\\
\cline{2-3}  \cline{5-6} 
$a_0$&7.15098    & (4.58221,  9.71976)    && 7.24832    & (4.57688,    9.91976)    \\
$a_1$&-0.04131   & (-0.06309,  -0.01953)   && -0.04193   & (-0.06439,   -0.01947)   \\
$a_2$&-0.03895   & (-0.06161,  -0.01628)   && -0.03963   & (-0.06311,   -0.01614)   \\
$b_0$&-1.61714   & (-3.88543,  0.65115)   && -1.54946   & (-3.92305,   0.82413)   \\
$b_1$&0.00854    & (-0.01489, 0.03196)    && 0.00811    & (-0.01605,   0.03227)   \\
$b_2$&0.01288    & (-0.01030, 0.03607)    && 0.01270    & (-0.01121,   0.03661)    \\
$R(25,35;60)$&0.95668    & (0.00153,  1.00000)    && 0.96179    & (0.00134,    1.00000)    \\
$R(35,50;60)$&0.56504    & (0.02025,  0.98790)    && 0.60006    & (0.02406,   0.98917)    \\
$E(25,35)$&125.15200 & (27.19831, 575.88220) && 133.67967 & (27.05276,  660.57060) \\ \hline
\end{tabular}
\end{table}

\section{Concluding Remarks \label{sec:CR}}

The aim of the present research was to develop robust statistical inference for  one-shot device testing model under lognormal distribution. We have presented a family of weighted minimum DPD estimators, as a generalization of the classical MLE. Based on this family, we are also able to develop confidence intervals and a family of Wald-type tests, which generalizes the classical Wald test. The robustness of the proposed statistics is evaluated through the study of the Influece Function and an extensive simulation study. Finally, the methods developed here are applied to two real-life examples for illustrative purpose. The evidence from this study suggests the use of these  methods as a more robust alternative to those based on the MLE, with an unavoidable loss of efficiency. 

This research has raised  some questions that need  further investigation. Although the study was limited to a binary response (failure or success of the device when being tested), it may be of interest to consider the ``competing risks'' scenario, in which we assume the products under study can experience any one of various possible causes of failure. This  will be a challenging and interesting problem for further consideration.

\clearpage
%\begin{comment}
\appendix
\section{Proof of Results \label{App}}
\subsection{Proof of Result \ref{th:KL_MLE}}
\begin{proof}
We have
\begin{align*}
\sum_{i=1}^{I}\frac{K_{i}}{K}d_{KL}(\widehat{\boldsymbol{p}}_{i},\boldsymbol{\pi }_{i}(\boldsymbol{\theta}))&=\sum_{i=1}^{I}\frac{n_{i}}{K}\log \left( \dfrac{\frac{n_{i}}{K_{i}}}{F_{\omega}(W_i;\boldsymbol{x}_i,\boldsymbol{\theta})}\right) +\frac{K_i-n_{i}}{K}\log \left( \dfrac{\frac{K_i-n_{i}}{K_{i}}}{R_{\omega}(W_i;\boldsymbol{x}_i,\boldsymbol{\theta})}\right) \\
& =c-\frac{1}{K}\sum_{i=1}^{I}\left\{ n_{i}\log \left( F_{\omega}(W_i;\boldsymbol{x}_i,\boldsymbol{\theta})\right)  +(K_{i}-n_{i})\log \left( R_{\omega}(W_i;\boldsymbol{x}_i,\boldsymbol{\theta})\right) \right\} \\
& =c-\frac{1}{K}\log \left( \prod_{i=1}^{I}F_{\omega}^{n_{i}}(W_i;\boldsymbol{x}_i,\boldsymbol{\theta})   R_{\omega}^{K_{i}-n_{i}}(W_i;\boldsymbol{x}_i,\boldsymbol{\theta})\right) \\
& =c-\frac{1}{K}\log  \mathcal{L}(\boldsymbol{\theta};\boldsymbol{z}) ,
\end{align*}
 where $c=\sum_{i=1}^{I}\frac{n_{i}}{K_i}\log \left( \frac{n_{i}}{K_{i}}\right) +\frac{K_i-n_{i}}{K_i}\log \left( \frac{K_i-n_{i}}{K_{i}}\right)$ does not depend on the parameter $\boldsymbol{\theta}$. Therefore, the maximization of the likelihood is equivalent to the minimization of the weighted Kullback-Leibler divergence measure between probability vectors in  (\ref{eq:Mult_emp_prob_vector}) and (\ref{eq:Mult_theo_prob_vector}).
\end{proof}

\subsection{Proof of Result \ref{th:lognormal_asymp}}
\begin{proof}
Let us denote
\begin{align*}
\boldsymbol{u}_{ij}(\boldsymbol{\theta})  & = \frac{\partial\log\pi_{ij}(\boldsymbol{\theta})}{\partial\boldsymbol{\theta}}  =\frac{1}{\pi_{ij}(\boldsymbol{\theta})}\left(\frac{\partial\pi_{ij}(\boldsymbol{\theta})}{\partial\boldsymbol{\theta}}\right)= \frac{1}{\pi_{ij}(\boldsymbol{\theta})}\boldsymbol{\delta}_i\boldsymbol{x}_i,
\end{align*}
with $\boldsymbol{\delta}_i$  as given in (\ref{eq:lognormal_Delta}). Upon using Theorem 3.1 of Ghosh and Basu (2013), we have%

\[
\sqrt{K}\left(  \widehat{\boldsymbol{\theta}}_{\beta}-\boldsymbol{\theta}^{0}\right)  \overset{\mathcal{L}}{\underset{K\mathcal{\rightarrow}\infty}{\longrightarrow}}\mathcal{N}\left(  \boldsymbol{0}_{2(J+1)},\boldsymbol{J}_{\beta}^{-1}(\boldsymbol{\theta}^{0})\boldsymbol{K}_{\beta}(\boldsymbol{\theta}^{0})\boldsymbol{J}_{\beta}^{-1}(\boldsymbol{\theta}^{0})\right)  ,
\]
where
\begin{align*}
\boldsymbol{J}_{\beta}(\boldsymbol{\theta}) &  =\sum_{i=1}^{I}\sum_{j=1}^{2}\frac{K_i}{K}\boldsymbol{u}_{ij}(\boldsymbol{\theta})\boldsymbol{u}_{ij}^{T}(\boldsymbol{\theta})\pi_{ij}^{\beta+1}(\boldsymbol{\theta}),\\
\boldsymbol{K}_{\beta}(\boldsymbol{\theta}) &  =\left(  \sum_{i=1}^{I}\sum_{j=1}^{2}\frac{K_i}{K}\boldsymbol{u}_{ij}(\boldsymbol{\theta})\boldsymbol{u}_{ij}^{T}(\boldsymbol{\theta})\pi_{ij}^{2\beta+1}(\boldsymbol{\theta})-\sum_{i=1}^{I}\frac{K_i}{K}\boldsymbol{\xi}_{i,\beta}(\boldsymbol{\theta})\boldsymbol{\xi}_{i,\beta}^{T}(\boldsymbol{\theta})\right)  ,
\end{align*}
with
\begin{align*}
\boldsymbol{\xi}_{i,\beta}(\boldsymbol{\theta})   =\sum_{j=1}^{2}\boldsymbol{u}_{ij}(\boldsymbol{\theta})\pi_{ij}^{\beta+1}(\boldsymbol{\theta}) =\boldsymbol{\delta}_i\boldsymbol{x}_{i}\sum_{j=1}^{2}(-1)^{j+1}\pi_{ij}^{\beta}(\boldsymbol{\theta}).
\end{align*}
Now,  we have \ $\boldsymbol{u}_{ij}(\boldsymbol{\theta})\boldsymbol{u}_{ij}^{T}(\boldsymbol{\theta})  =\frac{1}{\pi_{ij}^{2}(\boldsymbol{\theta})}\boldsymbol{\Delta}_{i}\boldsymbol{x}_i \boldsymbol{x}_i^T$, with  $\boldsymbol{\Delta}_{i}=\boldsymbol{\delta}_{i}\boldsymbol{\delta}^T_{i}$. It then follows that%
\begin{align*}
\boldsymbol{J}_{\beta}(\boldsymbol{\theta}) =\sum_{i=1}^{I}\frac{K_i}{K}\boldsymbol{\Delta}_{i}\sum_{j=1}^{2}\pi_{ij}^{\beta-1}(\boldsymbol{\theta}) =\sum_{i=1}^{I}\frac{K_i}{K}\boldsymbol{\Delta}_{i}\left(  \pi_{i1}^{\beta-1}(\boldsymbol{\theta})+\pi_{i2}^{\beta-1}(\boldsymbol{\theta})\right)\boldsymbol{x}_i \boldsymbol{x}_i^T.
\end{align*}
In a similar manner, $\boldsymbol{\xi}_{i,\beta}(\boldsymbol{\theta})\boldsymbol{\xi}_{i,\beta}^{T}(\boldsymbol{\theta})=\boldsymbol{\Delta}_{i}\left(  \sum_{j=1}^{2}(-1)^{j+1}\pi_{ij}^{\beta}(\boldsymbol{\theta})\right)  ^{2}$\ and%
\[
\boldsymbol{K}_{\beta}(\boldsymbol{\theta})=\sum_{i=1}^{I}\frac{K_i}{K}\boldsymbol{\Delta}_{i}\left(  \sum_{j=1}^{2}\pi_{ij}^{2\beta-1}(\boldsymbol{\theta})-\left(  \sum_{j=1}^{2}(-1)^{j+1}\pi_{ij}^{\beta}(\boldsymbol{\theta})\right)  ^{2}\right)\boldsymbol{x}_i \boldsymbol{x}_i^T  .
\]
Because 
\[
\sum_{j=1}^{2}\pi_{ij}^{2\beta-1}(\boldsymbol{\theta})-\left(  \sum_{j=1}^{2}(-1)^{j+1}\pi_{ij}^{\beta}(\boldsymbol{\theta})\right)  ^{2}=\pi_{i1}(\boldsymbol{\theta})\pi_{i2}(\boldsymbol{\theta})\left(  \pi_{i1}^{\beta-1}(\boldsymbol{\theta})+\pi_{i2}^{\beta-1}(\boldsymbol{\theta})\right)  ^{2},
\]
we have
\[
\boldsymbol{K}_{\beta}(\boldsymbol{\theta})=\sum_{i=1}^{I}\frac{K_i}{K}\boldsymbol{\Delta}_{i}\pi_{i1}(\boldsymbol{\theta})\pi_{i2}(\boldsymbol{\theta})\left(  \pi_{i1}^{\beta-1}(\boldsymbol{\theta})+\pi_{i2}^{\beta-1}(\boldsymbol{\theta})\right)  ^{2} \boldsymbol{x}_i \boldsymbol{x}_i^T.
\]
\end{proof}
%\end{comment}

\subsection{Proof of Result \ref{th:asymp_test}}

\begin{proof}
Let $\boldsymbol{\theta}^{0}\in \Theta_0 $ be the true value of parameter $\boldsymbol{\theta}$. It is then clear that 

\begin{align*}
\boldsymbol{m}( \widehat{\boldsymbol{\theta}}_{\beta }) =\boldsymbol{m}( \boldsymbol{\theta}^{0}) +\boldsymbol{M}^{T}( \widehat{\boldsymbol{\theta}}_{\beta }) ( \widehat{\boldsymbol{\theta}}_{\beta }-\boldsymbol{\theta}^{0}) +o_{p}\left( \left\Vert \widehat{\boldsymbol{\theta}}_{\beta }-\boldsymbol{\theta}^{0}\right\Vert \right) =\boldsymbol{M}^{T}( \widehat{\boldsymbol{\theta}}_{\beta }) ( \widehat{\boldsymbol{\theta}}_{\beta }-\boldsymbol{\theta}^{0}) +o_{p}\left(K^{-1/2}\right) .
\end{align*}
But, $\sqrt{K}\left( \widehat{\boldsymbol{\theta}}_{\beta }-\boldsymbol{\theta}^{0}\right) \underset{K\rightarrow \infty }{\overset{\mathcal{L}}{\longrightarrow }}\mathcal{N}\left( \boldsymbol{0}_{J+1},\boldsymbol{\Sigma }_{\beta }( \widehat{\boldsymbol{\theta}}_{\beta }) \right)$. Therefore, we have 

\begin{equation*}
\sqrt{K}\boldsymbol{m}( \widehat{\boldsymbol{\theta}}_{\beta }) \underset{K\rightarrow \infty }{\overset{\mathcal{L}}{\longrightarrow }}\mathcal{N}\left( \boldsymbol{0}_{r},\boldsymbol{M}^{T}( \boldsymbol{\theta}^{0}) \boldsymbol{\Sigma }_{\beta }(\boldsymbol{\theta}^0) \boldsymbol{M}( \boldsymbol{\theta}^{0}) \right)
\end{equation*}
and taking into account that $rank(\boldsymbol{M}\left( \boldsymbol{\theta}^{0}\right) )=r$, we readily obtain

\begin{equation*}
K\boldsymbol{m}^{T}( \widehat{\boldsymbol{\theta}}_{\beta }) \left( \boldsymbol{M}^{T}( \boldsymbol{\theta}^{0}) \boldsymbol{\Sigma }_{\beta }( \boldsymbol{\theta}^{0}) \boldsymbol{M}( \boldsymbol{\theta}^{0}) \right) ^{-1}\boldsymbol{m}( \widehat{\boldsymbol{\theta}}_{\beta }) \underset{K\rightarrow \infty }{\overset{\mathcal{L}}{\longrightarrow }}\chi _{r}^{2}.
\end{equation*}%

As $\left( \boldsymbol{M}^{T}( \widehat{\boldsymbol{\theta}}_{\beta })\boldsymbol{\Sigma }_{\beta }( \widehat{\boldsymbol{\theta}}_{\beta}) \boldsymbol{M}( \widehat{\boldsymbol{\theta}}_{\beta })\right) ^{-1}$ is a consistent estimator of $\left( \boldsymbol{M}^{T}( \boldsymbol{\theta}^{0}) \boldsymbol{\Sigma }_{\beta }( \boldsymbol{\theta}^{0}) \boldsymbol{M}( \boldsymbol{\theta}^{0})\right) ^{-1}$, we get 
\begin{equation*}
W_{K}( \widehat{\boldsymbol{\theta}}_{\beta }) \underset{K\rightarrow\infty }{\overset{\mathcal{L}}{\longrightarrow }}\chi _{r}^{2}.
\end{equation*}
\end{proof}

\end{document}